\documentclass[prl, superscriptaddress, twocolumn, nobibnotes, aps]{revtex4-2}

\usepackage{stix}

\usepackage{comment}
\usepackage[dvipsnames]{xcolor}
\usepackage{hyperref}
\hypersetup{
    colorlinks=true,
    linkcolor=BlueViolet,
    citecolor=Blue,
    urlcolor=BlueViolet,
}
\usepackage{graphicx}
\usepackage{amsmath,amsfonts}
\usepackage{mathtools}
\usepackage{bm}
\usepackage{textcomp}
\usepackage[export]{adjustbox}
\usepackage[capitalize]{cleveref}
\usepackage{braket}

\newcommand{\bvec}[1]{\boldsymbol{#1}}
\newcommand{\dd}{\mathrm{d}}

\newcommand{\sout}[1]{{\color{OrangeRed}#1}}

\renewcommand{\sout}[1]{}

\usepackage{tikz}
\makeatletter
\def\parsecomma#1,#2\endparsecomma{\def\page@x{#1}\def\page@y{#2}}
\tikzdeclarecoordinatesystem{page}{
    \parsecomma#1\endparsecomma
    \pgfpointanchor{current page}{north east}
    \pgf@xc=\pgf@x%
    \pgf@yc=\pgf@y%
    \pgfpointanchor{current page}{south west}
    \pgf@xb=\pgf@x%
    \pgf@yb=\pgf@y%
    \pgfmathparse{(\pgf@xc-\pgf@xb)/2.*\page@x+(\pgf@xc+\pgf@xb)/2.}
    \expandafter\pgf@x\expandafter=\pgfmathresult pt
    \pgfmathparse{(\pgf@yc-\pgf@yb)/2.*\page@y+(\pgf@yc+\pgf@yb)/2.}
    \expandafter\pgf@y\expandafter=\pgfmathresult pt
}
\makeatother
\newenvironment{supplement}{\clearpage\pagenumbering{gobble}}{}
\newcommand{\supppage}[1]{%
  \begin{tikzpicture}[remember picture,overlay,inner sep=0pt,outer sep=0pt]
    \node[anchor=north west] at (page cs:-1,1) {\includegraphics[page=#1]{supp.pdf}};
  \end{tikzpicture}
  \clearpage
}

\newcommand{\arxivSubmit}{%
\begin{supplement}
  \supppage{1}
  \supppage{2}
  \supppage{3}
  \supppage{4}
  \supppage{5}
  \supppage{6}
  \supppage{7}
  \supppage{8}
\end{supplement}
}

\begin{document}
\title{Supercell Wannier functions and a faithful low-energy model for Bernal bilayer graphene}
\author{Ammon Fischer}
\thanks{These authors contributed equally.}
\email{ammon.fischer@rwth-aachen.de}
\affiliation{Institute for Theory of Statistical Physics, RWTH Aachen University, and JARA Fundamentals of Future Information Technology, 52062 Aachen, Germany}

\author{Lennart Klebl}
\thanks{These authors contributed equally.}
\affiliation{I. Institute for Theoretical Physics, Universität Hamburg, Notkestraße 9-11, 22607 Hamburg, Germany}

\author{Dante M.~Kennes}
\affiliation{Institute for Theory of Statistical Physics, RWTH Aachen University, and JARA Fundamentals of Future Information Technology, 52062 Aachen, Germany}
\affiliation{Max Planck Institute for the Structure and Dynamics of Matter, Center for Free Electron Laser Science, 22761 Hamburg, Germany}
\author{Tim O.~Wehling}
\affiliation{I. Institute for Theoretical Physics, Universität Hamburg, Notkestraße 9-11, 22607 Hamburg, Germany}
\affiliation{The Hamburg Centre for Ultrafast Imaging, 22761 Hamburg, Germany}

\date{\today}
\begin{abstract}
We derive a minimal low-energy model for Bernal bilayer graphene and related rhombohedral graphene multilayers at low electronic densities by constructing Wannier orbitals defined in real-space supercells of the original primitive cell. 
Starting from an \textit{ab-initio} electronic structure theory comprising the atomic carbon $p_z$-orbitals, momentum locality of the Fermi surface pockets around $K,K'$ is circumvented by backfolding the $\pi$-bands to the concomitant mini-Brillouin zone of the supercell, reminiscent of their (twisted) moiré counterparts.
The supercell Wannier functions reproduce the spectral weight and Berry curvature of the microscopic model and offer an intuitive real-space picture of the emergent physics at low electronic densities being shaped by flavor-polarized wave packets with mesoscopic extent.
By projecting an orbital-resolved, dual-gated Coulomb interaction to the effective Wannier basis, we find that the low-energy physics of Bernal bilayer graphene is governed by weak electron-electron interactions. 
Our study bridges between existing continuum theories and \textit{ab-initio} studies of small Fermi pocket systems like rhombohedral graphene stacks by providing a symmetric lattice description of their low-energy physics. 
\end{abstract}

\maketitle

\paragraph{Introduction.}

\begin{figure}
    \centering
    \includegraphics[width=\columnwidth]{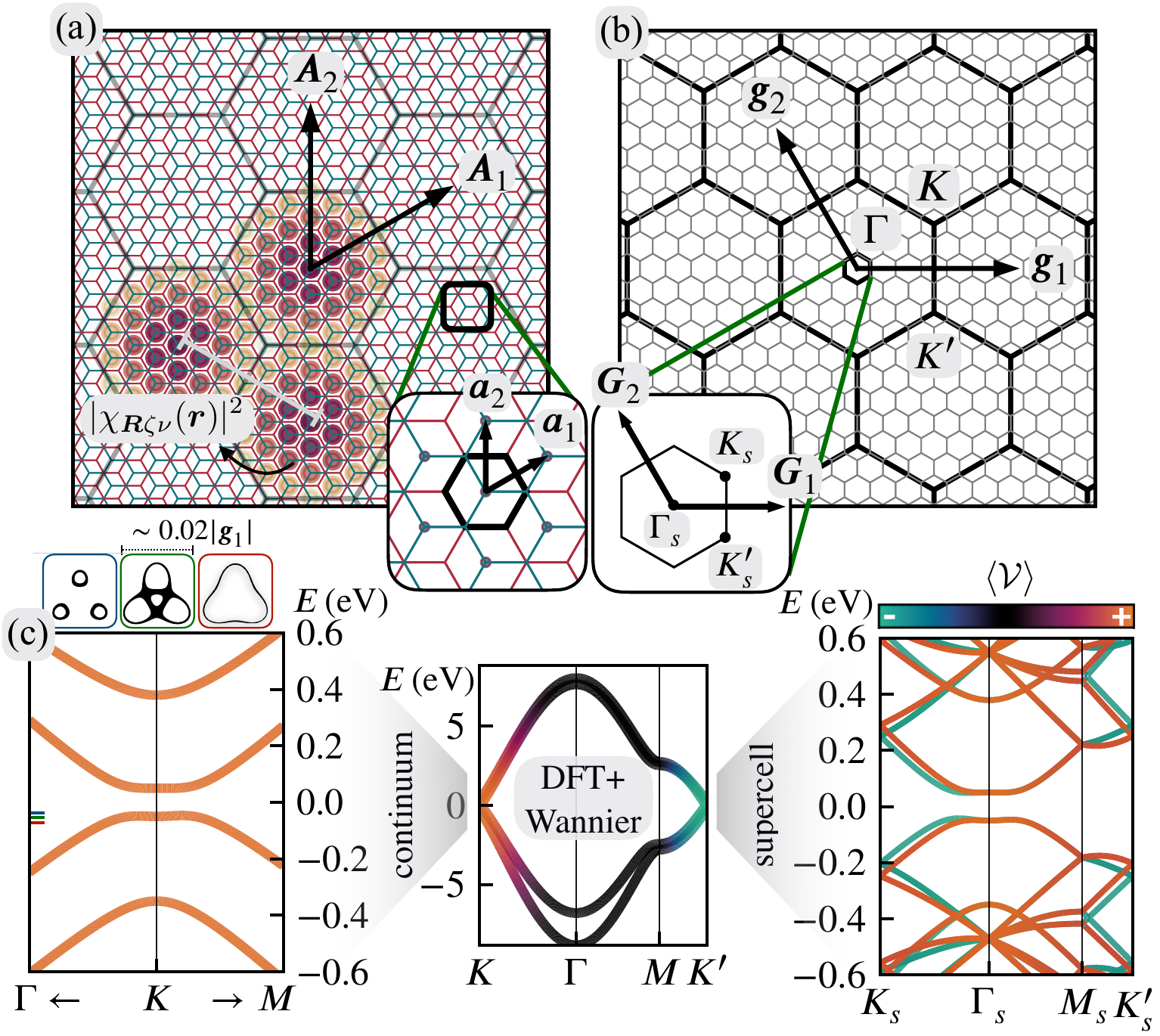}
    \caption{Supercell Wannier construction in Bernal bilayer graphene (BBG). Real-space (a) and momentum space (b) geometry of the BBG supercell and mini Brillouin zone (BZ). The effective Wannier functions $|\chi_{\zeta \nu}(\bvec r)|^2$ of the low-energy modes are localized on length scales associated to the supercell lattice constant $L_s = |\bvec A_{i}| = n_s |\bvec a_i|$.
    The original BZ is folded into the mini-BZ spanned by reciprocal lattice vectors $\bvec G_i = \bvec g_i / n_s$.
    The left inset shows the primitive real-space unit cell and the density of a Bloch wave packet $\psi_K(\bvec r)$. The right inset indicates the mini-BZ with its high symmetry points. (c) Bandstructure obtained from DFT \& Wannierization of the carbon $p_z$-orbitals in the presence of an interlayer potential difference of $\Delta = 100$ meV. Continuum limit and downfolded bands in the effective supercell are indicated as well as the Fermi surfaces across the Lifshitz transition at low densities (black markers). The bands are color-coded according to their valley expectation value $\langle \mathcal{V} \rangle$.}
    \label{fig:fig1}
\end{figure}

Since the initial experimental observation of correlated metallic and insulating states as well as superconductivity in Bernal bilayer graphene (BBG)~\cite{seiler2022quantum,zhou2022isospin}, numerous experimental~\cite{holleis2023ising, zhou2022isospin,zhang2023enhanced, winterer2023ferroelectric,seiler2023interactiondriven,seiler2024layer,li2024tunable,tsui2024direct,xie2024optical,seiler2024layerselective} and theoretical efforts~\cite{chatterjee2022inter,szabo2022competing,szabo2022metals,PhysRevB.105.134524,li2023charge,pantaleon2022superconductivity,pantaleon2023superconductivity,cea2022superconductivity,jimeno2023superconductivity,PhysRevB.107.L041111,dai2022quantum,ghazaryan2023multilayer,wei2023,wagner2023superconductivity,fischer2023spin,zhumagulov2023emergent} have been undertaken to unravel the microscopic origin and interplay between various symmetry-broken states at low electronic densities. Experiments have shown the existence of ferromagnetic phases in the form of half- and quarter metals~\cite{zhou2021half,seiler2022quantum}, anomalous hall crystals~\cite{seiler2022quantum}, (generalized) Wigner crystals~\cite{seiler2022quantum,tsui2024direct} and superconductivity in BBG~\cite{zhou2022isospin,zhang2023enhanced,holleis2023ising,li2024tunable} and rhombohedral trilayer graphene~\cite{zhou2021superconductivity} subject to an external displacement field. Opposite to the formation of true electronic flat bands in their twisted counterparts~\cite{cao_unconventional_2018,yankowitz2019tuning,lu2019superconductors,saito2020independent,stepanov2020untying,oh2021evidence,cao2021nematicity,zhang2021ascendance, park2021tunable,cao2021pauli,kim2022evidence,liu2022isospin,kerelsky2019maximized,Balents2020}, pristine multilayer graphene stacks host small regions in momentum space where the electronic bands are flattened upon an electrical field induced lifting of the inversion-symmetry protected Dirac points $K^{\nu}$~\cite{jung2014accurate,rozhkov2016electronic,pantaleon2023superconductivity}. This leads to the formation of van-Hove singularities on the electron and hole-doped side~\cite{seiler2024layer,li2024tunable}, which stabilize symmetry-broken phases mediated by electron-electron interactions~\cite{seiler2022quantum,zhou2021superconductivity}. However, the low-energy bands of multilayer graphene are neither flat in the entire Brillouin zone, nor separated from the remaining valence (conduction) bands by a well-defined energy gap. Instead, the Fermi surface near charge neutrality consists of highly localized pockets around the valleys $K^{\nu}$ as shown in \cref{fig:fig1}~(c).

The strategy that is commonly applied to describe the low-energy physics in BBG and related rhombohedral graphene stacks is to resort to a full momentum-space formalism by deriving a $\bvec k \cdot \bvec p$ expansion of the Hamiltonian around the non-equivalent valleys $K^{\nu}$~\cite{jung2014accurate,rozhkov2016electronic} instead of a symmetric lattice description. While this description is sufficient when modeling electronic properties, including the fermiology and bandstructure of various multilayer graphene stacks in the presence of external fields and proximity-induced spin-orbit coupling~\cite{gmitra2009band,gmitra2015graphene,zollner2016theory,sierra2021van,seiler2024layerselective}, the continuum formulation unavoidably leaves the realm of \textit{ab-initio} theories as soon as electron-electron interactions are added to the puzzle.
Truncating the model in momentum space invalidates the representation of local interaction terms and therefore hampers its applicability to well-established many-body techniques---most of which require a lattice description of the system under scrutiny including the functional renormalization group~\cite{metzner2012functional,Platt2013,Honerkamp2008} (FRG), dynamical mean-field theory~\cite{georges1996dynamical}, or tensor network based approaches~\cite{schollwock2011density,orus2014practical}. The central prerequisite to first-principle motivated modeling of correlated many-body physics in small Fermi pocket systems like multilayer graphene stacks~\cite{seiler2022quantum,zhou2021superconductivity,banszerus2023particle,Wirth2022Oct,atri2023spontaneous} or AlAs quantum wells~\cite{hossain2021,shkolnikov2002}, is therefore the choice of a local (real-space) basis capable to account for the atomic-scale and long-wavelength physics inherent to aforementioned material candidates with nearly free electron gas behavior at low densities. 

In the present manuscript, we construct a fully symmetric lattice model that
captures the low-energy physics of BBG and circumvents above obstacles. %
We reformulate the (interacting) physics of BBG and related rhombohedral
graphene stacks at low electronic densities in terms of supercell Wannier
functions (SWFs) that are localized on intermediate length scales $a_0 < L_s <
1/|\bvec q_{\mathrm{FS}}| \sim 60 a_0$ (with $|\bvec q_\mathrm{FS}|$ the
characteristic extent of the Fermi pockets). By comprising several hundreds of carbon
atoms on the atomic carbon-carbon scale $a_0$, the SWFs represent a
coarse-grained description of rhombohedral graphene stacks that links the atomic
representation of the electronic structure as obtained from first-principles,
e.g. via density functional theory (DFT) \& Wannierization of the carbon
$p_z$-orbitals~\cite{jung2014accurate} with a mesoscopic description at
intermediate length scales $L_s$. Therefore, the SWFs yield an elegant
real-space analogue of the full momentum-space formalism adopted in the
continuum limit, where the locality and the effective length scales of emergent
phenomena are hidden. %
We demonstrate that spectral features and the local Berry curvature 
of the valley-local flat bands near $K^{\nu}$ can be faithfully represented by two $p_{\pm}$-orbitals per valley that span the $U(4) \otimes U(4)$ symmetry space of spin, valley and layer degrees of freedom. 
This allows to downfold realistic (dual-gated) Coulomb interactions to the effective Wannier basis including orbitally resolved short- and long-ranged interaction components~\cite{Wehling2011,rosner2015wannier}. 
Our work not only demonstrates that the low-energy physics in BBG is governed by weak electron-electron interactions, but at the same time provides a natural real-space description that bridges to the emergent physics in twisted graphene multilayers.

\paragraph*{Supercell construction for small Fermi pocket systems.}

To circumvent a full momentum-space formalism of the low-energy physics in BBG and related small Fermi pocket systems, we propose a real-space supercell construction inspired by the moiré pattern in twisted (graphene) multilayers~\cite{bistritzer2011moire,dos2007graphene, shallcross2010electronic, morell2010flat}, albeit without a twist between adjacent layers.
For the former, Wannier models have become state of the art for studies going beyond the single-particle picture~\cite{carr2019relaxation, carr2019wannier, song2022magic, hu2023symmetric, yu2023magic, rai2023dynamicalcorrelationsordermagicangle, bennett2024twisted}.
Instead of describing the crystal structure of BBG in terms of its primitive
unit cell with Bravais lattice vectors $\bvec a_1 = (\sqrt{3}/2,
1/2)^{\mathrm{T}}$ and $\bvec a_2 = (0,1)^{\mathrm{T}}$, we set up a
superlattice with rescaled lattice vectors $\bvec A_i = n_s \bvec
a_i$~\footnote{In principle, the construction works for all supercells that are
the result of a linear integer map applied to the primitive lattice vectors,
i.e., $(\bvec A_1, \bvec A_2)^T = \mathcal N\cdot (\bvec a_1, \bvec a_2)^T$,
with $\mathcal N$ a $2\times 2$ matrix with integer entries.}, where $n_s$
denotes the linear scaling factor of the supercell. As shown in
Fig.~\ref{fig:fig1}~(a), the superlattice contains $n_s^2$ primitive unit cells
on the atomic scale and is characterized by the mesoscopic length scale $L_s =
|\bvec A_i| = n_s |\bvec a_i|$. The associated Brillouin zone (BZ) is shrunk to
a mini-BZ recovering momentum-space periodicity with respect to the reciprocal
lattice vectors $\bvec G_i = \bvec g_i/n_s$ as indicated in
Fig.~\ref{fig:fig1}~(b). The high-symmetry points of the mini-BZ are labeled
according to their counterparts in the primitive BZ. Depending on the choice of
the scaling factor $n_s$, the original graphene valleys $K^{\nu}$, $\nu = \pm $
are mapped to either the $\Gamma_s$-point [$\mathrm{mod} (n_s, 3) = 0$] as shown
in~\cref{fig:fig1}~(b), or onto the supercell $K_{s}^{\nu}$-points
[$\mathrm{mod}(n_s, 3) \neq 0$].
To construct the Hamiltonian of the supercell from first-principles, we
first consider an \textit{ab-initio} Hamiltonian of BBG formulated in the basis
of the four carbon $p_z$-orbitals centered at the sublattices $X=(A_1, B_1, A_2,
B_2)$ as routinely derived from DFT \& Wannierization~\cite{jung2014accurate}.
As the supercell construction itself preserves translational invariance on the
carbon-carbon bond scale, the Hamiltonian associated to the primitive cell can
be mapped to the supercell geometry. Hence we avoid costly DFT simulations of
all $4n_s^2$ carbon $p_z$-orbitals that reside within the supercell, see
Supplementary Material (SM)~\cite{SM} for technical details.
The symmetries of the supercell Hamiltonian
$\mathcal G^0$ are carried over from the microscopic model, i.e.,
$\mathcal{G}^0=C_{3v} \otimes SU(2) \otimes \mathcal{T}$. Here, $C_{3v}$ is the
point group of the superlattice that contains a three-fold rotation around the
$z$-axis and three vertical mirror planes, the $SU(2)$-symmetry acts in spin
space and $\mathcal{T}$ denotes time-reversal symmetry. At low electronic
densities, the system further possesses an approximate valley conservation
$U_{\nu}(1)$ that is reflected in an orbital-dependent phase factor $e^{i \bvec
K^{\nu} \bvec r}$ of the respective Bloch states. In the supercell picture, the
valley character of the Bloch states $|\psi_{\bvec k b} \rangle $ can be
disentangled by evaluating expectation values $\langle  \mathcal{V} \rangle =
\langle \psi_{\bvec k b} |  \mathcal{V}| \psi_{\bvec k b} \rangle$ of a
Haldane-like operator
$\mathcal{V}$~\cite{haldane1988model,ramires2018electrically}, see SM~\cite{SM}.
As demonstrated in \cref{fig:fig1}~(c), the back-folded bands in the mini-BZ of
the supercell originating from valley $K^{\nu}$ indeed carry a well-defined
expectation value $\langle \mathcal{V} \rangle \sim \nu$, whereas remote energy
states originating from states near $\Gamma$ show no distinct valley
polarization. 

\begin{figure*}[t!]
    \centering
    \includegraphics[width=0.9\textwidth]{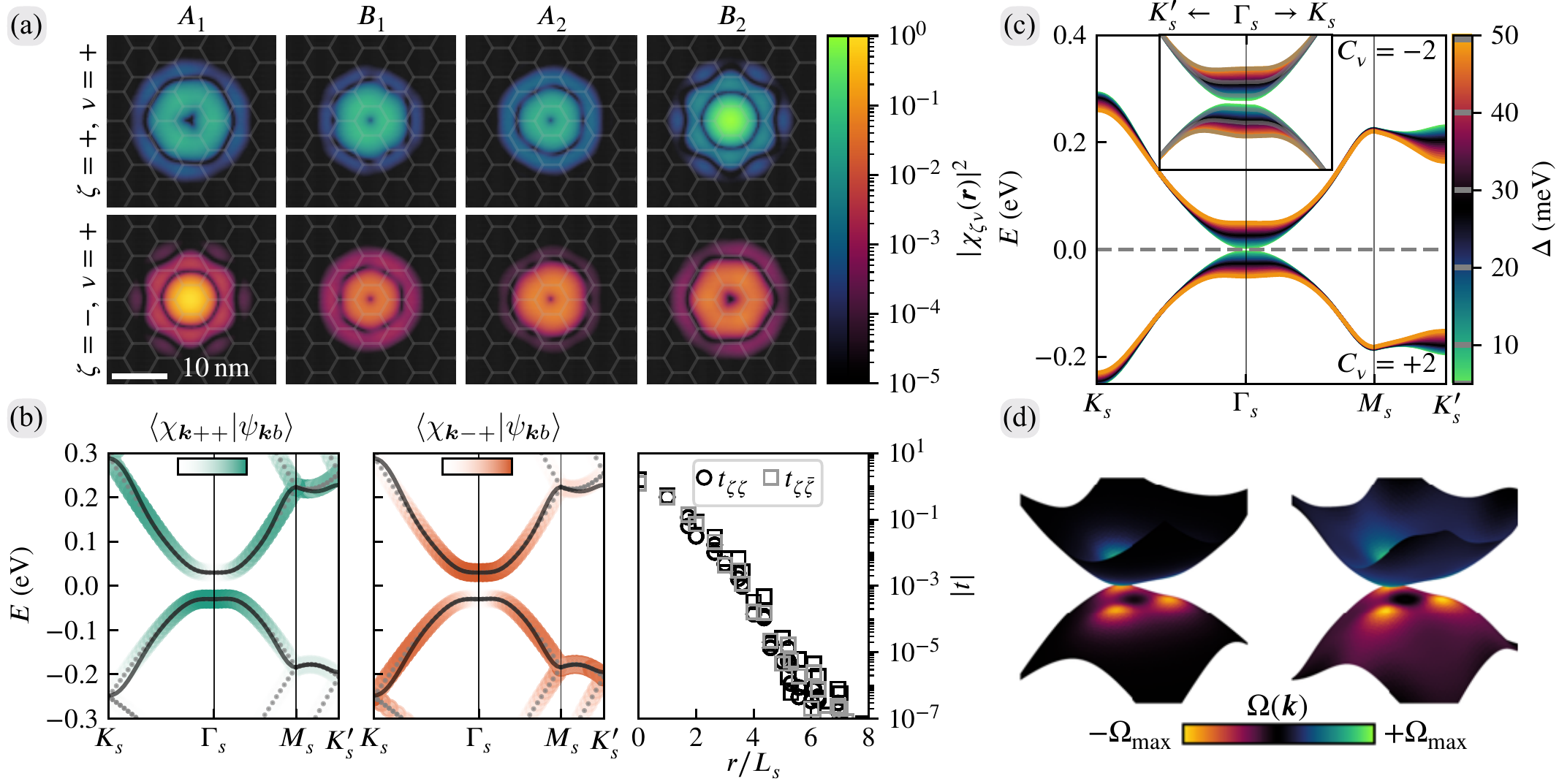}
    \caption{Supercell Wannier orbitals and minimal low-energy model for Bernal bilayer graphene. (a) Probability density of the supercell Wannier functions (rows) on each of the four BBG sites (columns, sublattice $A, B$ and layer $1,2$). The size of the supercell is $n_s=24$, which amounts to a superlattice constant of $|\bvec A_{1,2}| = 5.9\,\mathrm{nm}$ (supercells are indicated as faint white lines). The interlayer potential difference is set to $\Delta = 60\,\mathrm{meV}$.
    (b) Projection of the Wannier orbitals on the original Bloch states of the supercell model. At the supercell Gamma point $\Gamma_s$ (that corresponds to the $K^\nu$ points of the primitive unit cell), the two orbitals carry the pure valley character of the valence/conduction band (left/center). The dispersion is accurately reproduced (black lines: Wannier model, dotted gray lines: supercell model) in a wide region around $\Gamma_s$. The right panel displays intra- and inter-orbital hopping parameters of the projected Hamiltonian that both decay exponentially as a function of distance $r$. 
    (c) Band structures for various displacement fields obtained from interpolating hopping parameters of the wannierized Hamiltonians for five grid points (gray markers in the colorbar). The inset shows how the interpolated Hamiltonians capture fine details of the band structure around $\Gamma_s$ (gray lines).
    (d) Berry curvature in the vicinity of $K^\nu$ in the atomic scale model of BBG (left) and the supercell Wannier model (right).}
    \label{fig:fig2}
\end{figure*}

The periodic supercell description of the atomic and electronic structure of BBG
allows to construct a low-energy model for BBG from \emph{first-principles} by
finding a minimal Wannier representation that captures the spectral weight and
Berry curvature of the valence and conduction bands near charge neutrality. In
this work, we focus on the intermediate regime $L_s < 1/|\bvec
q_{\mathrm{FS}}|$, where $|\bvec q_{\mathrm{FS}}|$ denotes the (inverse) linear
extent of the Fermi surface pockets around $K^{\nu}$. In this regime, only one
band per valley contributes to the Fermi surface close to the VHSs and the
artificial length scale set by the supercell does hence not infer with emergent
length scales associated to possible electronic ordering, see Fig.~1 in
the SM~\cite{SM}. To construct localized SWFs $\chi_{\bvec R \zeta \nu}(\bvec
r)$ in the presence of finite displacement fields, we apply the single-shot
projection technique~\cite{marzari1997maximally,Kang2019strong,carr2019wannier},
see SM~\cite{SM} for technical details.
Choosing $\mathrm{mod}(n_s, 3) = 0$ allows trial wave functions $g_{\bvec R \zeta \nu}(\bvec r)$ to be trivially symmetric (i.e., without phase factors) under the model's point group symmetries when sampled from Bloch states at $\Gamma_s$, i.e. $g_{\bvec R \zeta \nu}(\bvec r) \sim \psi_{\Gamma_s \zeta \nu}(\bvec r)$.
States originating from the valence (conduction) band manifold are polarized on the non-dimer orbitals $B_2$ $(A_1)$ as labeled by $\zeta = \pm$. Due to the emergent valley conservation $U_{\nu}(1)$ and the presence of time-reversal symmetry $\mathcal{T}$ and spin-$SU(2)$ symmetry, each state at $\Gamma_s$ is fourfold degenerate and we further separate Bloch states according to their valley polarization $\nu=\pm$. The trial orbitals $g_{\bvec R \zeta \nu}(\bvec r)$ transform as eigenstates of $C_{3v}$ with eigenvalue $w^{\zeta \nu}$, where $w = e^{i 2 \pi /3}$, and therefore resemble $p_{\pm} = p_x \pm i p_y$-orbitals with angular momentum $L_z = \nu \zeta$.
\paragraph*{Supercell Wannier functions.}

In \cref{fig:fig2}~(a) we show the valley-polarized SWFs $\chi_{\bvec R \zeta
+}(\bvec r)$ obtained in a BBG supercell with scaling factor $n_s=24$ and
interlayer potential difference of $\Delta=60$ meV. However, we stress that the
SWF construction presented explicitly for BBG can be generalized to arbitrary
rhombohedral graphene stacks, see SM~\cite{SM} for a discussion on monolayer and
ABC trilayer graphene. The SWFs are decomposed into contributions from the
microscopic sublattice- and layer degrees of freedom $X$: $\chi_{\bvec R \zeta
\nu}(\bvec r) = \{ \chi^{X}_{\bvec R \zeta \nu}(\bvec r) \}$. The amplitude of
the SWF with quantum number $\zeta$ is enhanced on the non-dimer orbital $B_2$
$(A_1)$ and decays exponentially on length scales associated to the BBG
supercell lattice constant $L_s$. To resolve the relative phase windings of the
SWFs on the individual sublattices, we separate contributions of the atomic
valley phase factor from the supercell envelope function $\chi_{\zeta \nu}(\bvec
r) = e^{i \bvec K_{\nu} \bvec r} W_{\zeta \nu}(\bvec r)$. The former features
$C_{3v}$ eigenvalues of $(w^{-1}, 1, 1, w)^{\nu}$ on the different sublattices,
whereas the supercell envelope function $W_{\zeta \nu}(\bvec r)$ has eigenvalues
$(w^{(\zeta+1)}, w^{\zeta}, w^{\zeta},w^{(\zeta-1)})^{\nu}$, see SM~\cite{SM}. Therefore, the total SWFs have the same $C_{3v}$ eigenvalue $w^{\zeta \nu}$ consistent with the $p_{\pm}$-orbital character of the trial functions. In particular, the sublattice components of the SWFs each have a relative phase winding of $2 \times 2 \pi$, which is reminiscent of the wave functions of chiral quasiparticles $\psi_{\bvec k \zeta \nu}(\bvec r) \sim e^{i \bvec k \bvec r} (1, \zeta e^{J i \nu \varphi_{\bvec k} })^{\mathrm{T}}$, where $\varphi_{\bvec k} = \arg(k_x + ik_y)$~\cite{mccann2006landau,rozhkov2016electronic} and $J=2$. The latter are well-known to govern the low-energy physics in the continuum formulation of rhombohedral multilayer graphene~\cite{mccann2006landau,rozhkov2016electronic}, where adiabatic propagation along a closed orbit yields a Berry phase of $J \pi$ for $J$-layered rhombohedral graphene.
This manifests in a half-integer valued valley Chern number $C_{\nu} = J \zeta \nu/2$~\cite{zhang2013valley,herzogarbeitman2023moire} characterized by an accumulation of Berry curvature $\Omega_{\nu}(\bvec k)$ in the vicinity of $K^{\nu}$. We demonstrate in \cref{fig:fig2}~(d) that the SWFs accurately reproduce the Berry curvature of the original Bloch states near charge neutrality.
Due to the momentum-space periodicity of the SWFs and the \emph{connectedness} of the lowest valence (conduction) bands, the Berry curvature of the original Hamiltonian can only be captured in a constrained energy window.
We find that the localized SWFs form a Haldane-like model within each valley featuring a valence (conduction) band valley Chern number $C_{\nu} =  \pm 2 \nu$ and thus represent an atomic limit in each valley. As time-reversal symmetry $\mathcal{T}$ connects different valley sectors $\chi_{\bvec R \zeta \bar \nu}(\bvec r) = \chi^*_{\bvec R \zeta \nu}(\bvec r)$, valence (conduction) bands of the full model retain Chern number $C=0$. 
In contrast to topologically (or chirally) obstructed low-energy models~\cite{herzogarbeitman2024topological}, the SWFs capture the entire spectral weight of the Bloch states near charge neutrality. The overlaps of the SWFs with the original Bloch states satisfy $\braket{\chi_{\bvec k\zeta\nu}|\psi_{\bvec kb}} = 1$ around $\Gamma_s$ as shown in \cref{fig:fig2}~(c), while decreasing at the BZ boundary.
Further evidence for the valley-polarized SWFs representing an atomic limit at low densities concerns the exponential decay of the hopping parameters $\mathsf{t}_{(\bvec R \zeta \nu) (\bvec R' \zeta' \nu' )}$ as function of distance, see the right panel of~\cref{fig:fig2}~(c).
Due to the explicit symmetries $U_{\nu}(1)$, $\mathcal{T}$ of the low-energy model, the hopping terms fulfill $\mathsf{t}_{(\bvec R \zeta \bar \nu) (\bvec R' \zeta' \bar \nu )} = [\mathsf{t}_{(\bvec R \zeta \nu) (\bvec R' \zeta' \nu)} ]^*$ and $\mathsf{t}_{(\bvec R \zeta \nu) (\bvec R' \zeta' \bar \nu )} = 0$,
such that the intra- and inter-orbital hopping terms in a single valley determine the full model. In~\cref{fig:fig2}~(c), we further demonstrate that the energetics of the valence (conduction) bands around charge neutrality are accurately reproduced by the SWFs. As shown in~\cref{fig:fig2}~(b), the effect of an external displacement field can be modeled by interpolating hopping terms at finite interlayer potential differences, while reproducing all characteristics of the low-energy bands including trigonal warping and the size of the single-particle band gap. 

\paragraph*{Coulomb interactions in the Wannier basis.}
\begin{figure}
    \centering
    \includegraphics[width=\columnwidth]{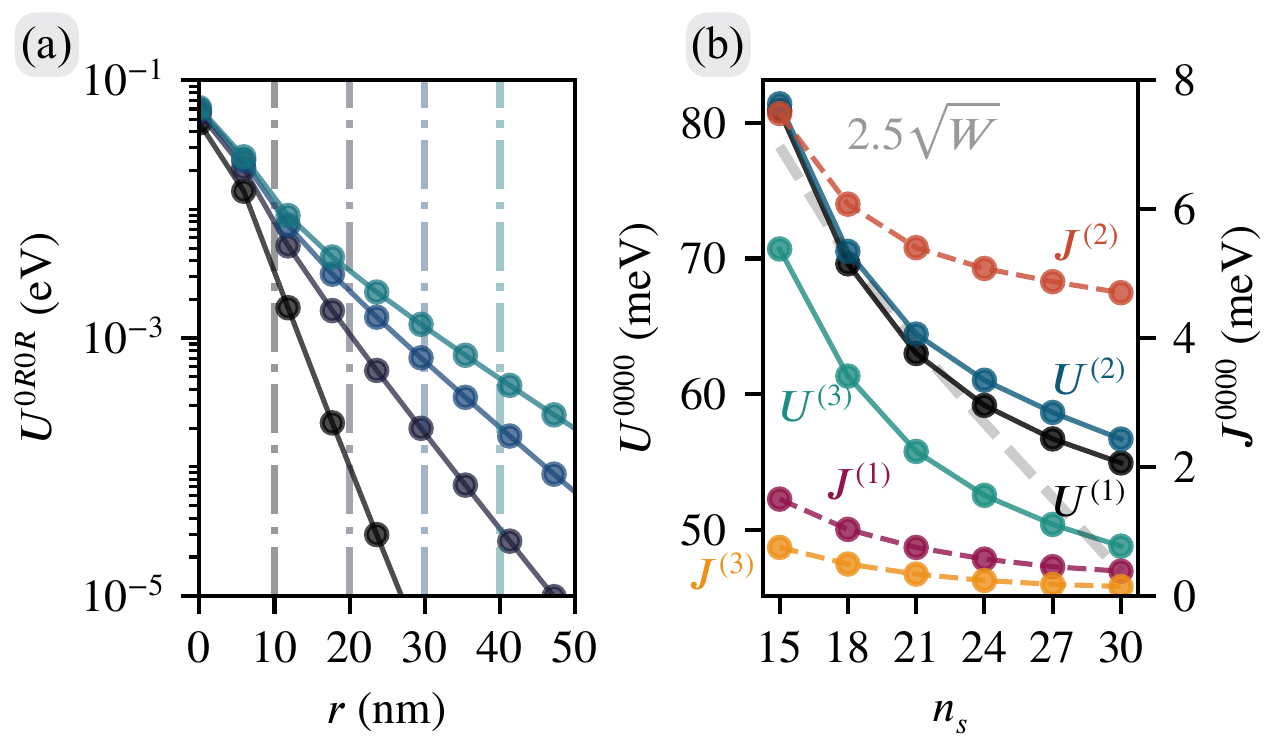}
    \caption{Strength of Coulomb interactions in the Wannier basis. (a) Density-density vertex element $U^{0R0R}$ as a function of distance $r$ for various gate distances (indicated as dash-dotted vertical lines of the respective color). The supercell size is chosen as $n_s=24$. (b) On-site Hubbard-Kanamori interaction components as a function of supercell size $n_s$. The left axis refers to Hubbard-$U$ terms and the right axis to Hund-$J$ couplings. The grey dashed line indicates the evolution of the bandwidth $W$ of the effective model.}
    \label{fig:fig3}
\end{figure}

Besides its major advantage of a symmetric lattice description, formulating the low-energy physics of BBG in terms of SWFs has the benefit of obtaining electron-electron interactions from first principles through the intrinsic connection to the atomistic basis of carbon $p_z$-orbitals.
We assume a realistic device
geometry~\cite{seiler2023interactiondriven,seiler2022quantum}, where a BBG sheet
is placed inbetween hBN spacer layers of thickness $\xi$ that determine the
dielectric environment $\epsilon\approx 4$~\cite{ghazaryan2023multilayer}. The
hBN/BBG/hBN stack is enclosed with metallic (graphite) gates. These assumptions
determine the overall strength $\alpha/\epsilon$ and characteristic distance
$\xi$ of the long-range part of the Coulomb interaction. The short-range
behavior of the Coulomb repulsion follows an Ohno form, with  parameters for
graphitic systems known from literature~\cite{Wehling2011,rosner2015wannier}.
This fixes the microscopic parametrization of the Coulomb interaction $V(r)$,
see SM~\cite{SM} for details. The supercell Wannier construction allows to project this first-principle $V(r)$ to the SWF basis (see \cref{fig:fig3}). First, we analyze density-density type interactions obtained by evaluating the Coulomb integral
\begin{equation}
U_{0 \bvec R 0 \bvec R}^{\alpha \alpha' \alpha \alpha'} =\! \iint \! \! \dd\bvec r\dd \bvec r' \,  |\chi_{0 \alpha} (\bvec r)|^2 |\chi_{\bvec R \alpha'} (\bvec r')|^2 V(|\bvec r - \bvec r'|) \,,
\label{eq:multilayer_coulomb_effective}
\end{equation}
for pairs of SWFs separated by $\bvec R$ in real-space. As demonstrated in \cref{fig:fig3}~(a), an exponential suppression sets in for distances larger than the gate distance $\xi=10\dots40\,\mathrm{nm}$. The short-ranged terms behave like $\sim1/r$ when $\xi$ is larger than the Wannier function spread. Second, we analyze the strength of local Hubbard-Kanamori interaction parameters
\begin{equation}
\begin{multlined}
J_{0000}^{\alpha_1 \dots \alpha_4} = \!\iint \!\! \dd\bvec r\dd \bvec r' \, \chi^*_{0 \alpha_1} (\bvec r)\chi^*_{0 \alpha_2} (\bvec r') \times{}
\\
\chi_{0 \alpha_4} (\bvec r')
\chi_{0 \alpha_3} (\bvec r)
V(|\bvec r - \bvec r'|) \,.
\end{multlined}
\label{eq:multilayer_hunds_effective}
\end{equation}
as well as their dependence on the supercell scaling factor $n_s$, see \cref{fig:fig3}~(b). In the enlarged pseudospin space including spin, valley and orbital (layer) degrees of freedom, the presence of time-reversal symmetry mandates four unique types of local interactions: (i) intra-orbital ($\zeta=\pm$ for $1,2$, respectively) $U^{(1,2)} = U^{(\zeta \nu)(\zeta \nu)(\zeta \nu)(\zeta \nu)}$ and inter-orbital $U^{(3)} = U^{(\zeta \nu)(\bar \zeta \nu)(\zeta \nu)(\bar \zeta \nu)}$ Hubbard interactions, (ii) inter-orbital, intra-valley coupling $J^{(1)} = J^{(\zeta \nu)(\bar \zeta \nu)(\bar \zeta \nu)(\zeta \nu)}$, (iii) intra-orbital, inter-valley coupling $J^{(2)} = J^{(\zeta \nu)(\zeta \bar \nu)(\zeta \bar \nu)(\zeta \nu)}$ and (iv) inter-orbital and inter-valley coupling $J^{(3)} = J^{(\zeta \nu)(\bar \zeta \bar \nu)(\bar \zeta \bar \nu)(\zeta \nu)}$. The local Hubbard interaction terms scale like $U^{(i)} \sim 1/n_s$ as expected from the Gaussian shape of the SWFs~\cite{song2022magic}. In particular, we find that for all supercell sizes $n_s=15\dots30$ considered within the scope of this work, the minimal model of BBG remains in the weakly-coupled regime $\max_{i}(U^{(i)}) < W$, where $W$ denotes the supercell bandwidth.
For the supercell scaling factor of $n_s=24$ used throughout the manuscript, we estimate the ratio of local interaction strength over bandwidth as $U/W \approx 0.15$.
For $W\gg\Delta$, the quadratic nature of the bands leads to the ratio of Hubbard interaction over band width scaling like $U/W\sim n_s$. The scaling behavior of $U/W$ changes for $W\gtrsim\Delta$, because the band width is not solely determined by the supercell scaling factor $n_s$ in this case.
Hund's coupling terms are suppressed by a factor of $\sim\!\!10$ compared to the intra-orbital and inter-orbital Hubbard interactions. The dominant Hund's interaction term in the SWF basis is the intra-orbital, inter-valley coupling $J^{(2)}$.

\paragraph{Discussion.}
In this Letter, we introduced SWFs as an efficient representation of the low-energy physics of BBG and related rhombohedral graphene stacks. The Wannier construction is based on the atomic-scale lattice of carbon atoms, but describes the emergent low-energy physics in terms of localized, flavor-polarized supercell orbitals that have the emergent symmetries of continuum models imprinted, i.e., valley-$U(1)$, spin-$SU(2)$ and time-reversal symmetry $\mathcal T$. To this end, we demonstrate that the spectral weight and Berry curvature of the low-energy Bloch states are faithfully represented by two valley-polarized $p_{\pm}$-orbitals located on the triangular sites of the superlattice. The local nature of our model implies that long-ranged Coulomb interactions can be cast into fairly local interactions, whose strength can be controlled by metallic gates. The values of the effective Coulomb interactions in the SWF basis are estimated from first principles, including density-density and exchange-type interactions like inter-valley Hund's couplings terms. 
The SFW construction remains valid as long as the superlattice length scale $L_s
= n_s |\bvec a_i| \lesssim 1 / |\bvec q_{\text{FS}}|$ does not exceed the length
scale associated to the width of the Fermi pockets $|\bvec q_\mathrm{FS}|$ at a
given electronic density. Otherwise, additional bands are back-folded to the
mini-BZ and need to be wannierized in order to sufficiently model the low-energy
physics. The `three-patch' continuum model~\cite{soejima2024anomalous} proposed
in the context of studying the translational symmetry-breaking anomalous Hall
crystal
phase~\cite{seiler2022quantum,herzogarbeitman2023moire,zeng2024gatetunable,soejima2024anomalous}
can in fact be understood as special case of our microscopic SWF construction
being valid in the intermediate regime $L_s \sim 1/ |\bvec q_{\text{FS}}|$,
where exactly three bands contribute to the Fermi surface. Since valley
conservation is included in the SWF model as an emergent symmetry, valley orders
are encoded as on-site order parameters, as opposed to the microscopic picture,
where the valley operator acts on bonds~\footnote{Using valley as a quantum
number in the supercell model amounts to convoluting the $\bvec q=0$ valley
operator with a $\bvec q=\bvec G_s$ charge density wave, see SM~\cite{SM}.}. This natural representation of valley as a quantum number in conjunction with the substantially reduced Hilbert space of the minimal SWF model enables the study of competing correlated phenomena in the phase diagram of rhombohedral graphene stacks with unbiased weak-coupling methods like FRG~\cite{metzner2012functional,Platt2013,diverge} from first-principles. To this end, it may be advisable to improve the downfolding scheme of the bare, dual-gated Coulomb interactions by using constrained random-phase approximation~\cite{aryasetiawan2004frequency} or FRG~\cite{kinza2015low-energy} schemes.
Due to its strength in representing first-principle interactions, the SWF model should allow to answer pressing questions regarding the formation of superconductivity from repulsive electron-electron or attractive phonon-induced interactions in various multilayer graphene stacks. Moreover, effects like proximity-induced Ising-SOC, microscopic defects and strain can readily be included on the level of the atomistic supercell, which allows for a material-realistic description relevant for experimental setups. In addition, the SWF procedure can be extended to other graphene multilayer stacks~\cite{Wirth2022Oct,fischer2023spin,atri2023spontaneous,garcia2023mixed,mcellistrim2023spectroscopic}, or modulation-doped AlAs quantum wells~\cite{hossain2021,shkolnikov2002} featuring valley-polarized and nematic behavior at low electronic densities.

\begin{acknowledgments}
We thank Matthew T.~Bunney, Francesco Grandi, and Jonas B.~Profe for fruitful discussions. This work was supported by the Excellence Initiative of the German federal and state governments, the Ministry of Innovation of North Rhine-Westphalia and the Deutsche Forschungsgemeinschaft (DFG, German Research Foundation). AF and DMK acknowledge funding by the DFG under RTG 1995, within the Priority Program SPP 2244 ``2DMP'' -- 443273985.
LK and TOW greatfully acknowledge support from the DFG through FOR 5249 (QUAST, Project No. 449872909) and SPP 2244 (Project No. 422707584).
TOW is supported by the Cluster of Excellence ``CUI: Advanced Imaging of Matter'' of the DFG (EXC 2056, Project ID 390715994).
DMK acknowledges support by the Max Planck-New York City Center for Nonequilibrium Quantum Phenomena. We acknowledge computational resources provided by RWTH Aachen University under project numbers rwth1420.
\end{acknowledgments}

\bibliography{references}

\begin{thebibliography}{97}%
\makeatletter
\providecommand \@ifxundefined [1]{%
 \@ifx{#1\undefined}
}%
\providecommand \@ifnum [1]{%
 \ifnum #1\expandafter \@firstoftwo
 \else \expandafter \@secondoftwo
 \fi
}%
\providecommand \@ifx [1]{%
 \ifx #1\expandafter \@firstoftwo
 \else \expandafter \@secondoftwo
 \fi
}%
\providecommand \natexlab [1]{#1}%
\providecommand \enquote  [1]{``#1''}%
\providecommand \bibnamefont  [1]{#1}%
\providecommand \bibfnamefont [1]{#1}%
\providecommand \citenamefont [1]{#1}%
\providecommand \href@noop [0]{\@secondoftwo}%
\providecommand \href [0]{\begingroup \@sanitize@url \@href}%
\providecommand \@href[1]{\@@startlink{#1}\@@href}%
\providecommand \@@href[1]{\endgroup#1\@@endlink}%
\providecommand \@sanitize@url [0]{\catcode `\\12\catcode `\$12\catcode
  `\&12\catcode `\#12\catcode `\^12\catcode `\_12\catcode `\%12\relax}%
\providecommand \@@startlink[1]{}%
\providecommand \@@endlink[0]{}%
\providecommand \url  [0]{\begingroup\@sanitize@url \@url }%
\providecommand \@url [1]{\endgroup\@href {#1}{\urlprefix }}%
\providecommand \urlprefix  [0]{URL }%
\providecommand \Eprint [0]{\href }%
\providecommand \doibase [0]{https://doi.org/}%
\providecommand \selectlanguage [0]{\@gobble}%
\providecommand \bibinfo  [0]{\@secondoftwo}%
\providecommand \bibfield  [0]{\@secondoftwo}%
\providecommand \translation [1]{[#1]}%
\providecommand \BibitemOpen [0]{}%
\providecommand \bibitemStop [0]{}%
\providecommand \bibitemNoStop [0]{.\EOS\space}%
\providecommand \EOS [0]{\spacefactor3000\relax}%
\providecommand \BibitemShut  [1]{\csname bibitem#1\endcsname}%
\let\auto@bib@innerbib\@empty
\bibitem [{\citenamefont {Seiler}\ \emph {et~al.}(2022)\citenamefont {Seiler},
  \citenamefont {Geisenhof}, \citenamefont {Winterer}, \citenamefont
  {Watanabe}, \citenamefont {Taniguchi}, \citenamefont {Xu}, \citenamefont
  {Zhang},\ and\ \citenamefont {Weitz}}]{seiler2022quantum}%
  \BibitemOpen
  \bibfield  {author} {\bibinfo {author} {\bibfnamefont {A.~M.}\ \bibnamefont
  {Seiler}}, \bibinfo {author} {\bibfnamefont {F.~R.}\ \bibnamefont
  {Geisenhof}}, \bibinfo {author} {\bibfnamefont {F.}~\bibnamefont {Winterer}},
  \bibinfo {author} {\bibfnamefont {K.}~\bibnamefont {Watanabe}}, \bibinfo
  {author} {\bibfnamefont {T.}~\bibnamefont {Taniguchi}}, \bibinfo {author}
  {\bibfnamefont {T.}~\bibnamefont {Xu}}, \bibinfo {author} {\bibfnamefont
  {F.}~\bibnamefont {Zhang}},\ and\ \bibinfo {author} {\bibfnamefont {R.~T.}\
  \bibnamefont {Weitz}},\ }\bibfield  {title} {\bibinfo {title} {Quantum
  cascade of correlated phases in trigonally warped bilayer graphene},\
  }\href@noop {} {\bibfield  {journal} {\bibinfo  {journal} {Nature}\ }\textbf
  {\bibinfo {volume} {608}},\ \bibinfo {pages} {298} (\bibinfo {year}
  {2022})}\BibitemShut {NoStop}%
\bibitem [{\citenamefont {Zhou}\ \emph {et~al.}(2022)\citenamefont {Zhou},
  \citenamefont {Holleis}, \citenamefont {Saito}, \citenamefont {Cohen},
  \citenamefont {Huynh}, \citenamefont {Patterson}, \citenamefont {Yang},
  \citenamefont {Taniguchi}, \citenamefont {Watanabe},\ and\ \citenamefont
  {Young}}]{zhou2022isospin}%
  \BibitemOpen
  \bibfield  {author} {\bibinfo {author} {\bibfnamefont {H.}~\bibnamefont
  {Zhou}}, \bibinfo {author} {\bibfnamefont {L.}~\bibnamefont {Holleis}},
  \bibinfo {author} {\bibfnamefont {Y.}~\bibnamefont {Saito}}, \bibinfo
  {author} {\bibfnamefont {L.}~\bibnamefont {Cohen}}, \bibinfo {author}
  {\bibfnamefont {W.}~\bibnamefont {Huynh}}, \bibinfo {author} {\bibfnamefont
  {C.~L.}\ \bibnamefont {Patterson}}, \bibinfo {author} {\bibfnamefont
  {F.}~\bibnamefont {Yang}}, \bibinfo {author} {\bibfnamefont {T.}~\bibnamefont
  {Taniguchi}}, \bibinfo {author} {\bibfnamefont {K.}~\bibnamefont
  {Watanabe}},\ and\ \bibinfo {author} {\bibfnamefont {A.~F.}\ \bibnamefont
  {Young}},\ }\bibfield  {title} {\bibinfo {title} {Isospin magnetism and
  spin-polarized superconductivity in bernal bilayer graphene},\ }\href@noop {}
  {\bibfield  {journal} {\bibinfo  {journal} {Science}\ }\textbf {\bibinfo
  {volume} {375}},\ \bibinfo {pages} {774} (\bibinfo {year}
  {2022})}\BibitemShut {NoStop}%
\bibitem [{\citenamefont {Holleis}\ \emph {et~al.}(2023)\citenamefont
  {Holleis}, \citenamefont {Patterson}, \citenamefont {Zhang}, \citenamefont
  {Yoo}, \citenamefont {Zhou}, \citenamefont {Taniguchi}, \citenamefont
  {Watanabe}, \citenamefont {Nadj-Perge},\ and\ \citenamefont
  {Young}}]{holleis2023ising}%
  \BibitemOpen
  \bibfield  {author} {\bibinfo {author} {\bibfnamefont {L.}~\bibnamefont
  {Holleis}}, \bibinfo {author} {\bibfnamefont {C.~L.}\ \bibnamefont
  {Patterson}}, \bibinfo {author} {\bibfnamefont {Y.}~\bibnamefont {Zhang}},
  \bibinfo {author} {\bibfnamefont {H.~M.}\ \bibnamefont {Yoo}}, \bibinfo
  {author} {\bibfnamefont {H.}~\bibnamefont {Zhou}}, \bibinfo {author}
  {\bibfnamefont {T.}~\bibnamefont {Taniguchi}}, \bibinfo {author}
  {\bibfnamefont {K.}~\bibnamefont {Watanabe}}, \bibinfo {author}
  {\bibfnamefont {S.}~\bibnamefont {Nadj-Perge}},\ and\ \bibinfo {author}
  {\bibfnamefont {A.~F.}\ \bibnamefont {Young}},\ }\href@noop {} {\bibinfo
  {title} {Ising superconductivity and nematicity in bernal bilayer graphene
  with strong spin orbit coupling}} (\bibinfo {year} {2023}),\ \Eprint
  {https://arxiv.org/abs/2303.00742} {arXiv:2303.00742 [cond-mat.supr-con]}
  \BibitemShut {NoStop}%
\bibitem [{\citenamefont {Zhang}\ \emph {et~al.}(2023)\citenamefont {Zhang},
  \citenamefont {Polski}, \citenamefont {Thomson}, \citenamefont
  {Lantagne-Hurtubise}, \citenamefont {Lewandowski}, \citenamefont {Zhou},
  \citenamefont {Watanabe}, \citenamefont {Taniguchi}, \citenamefont {Alicea},\
  and\ \citenamefont {Nadj-Perge}}]{zhang2023enhanced}%
  \BibitemOpen
  \bibfield  {author} {\bibinfo {author} {\bibfnamefont {Y.}~\bibnamefont
  {Zhang}}, \bibinfo {author} {\bibfnamefont {R.}~\bibnamefont {Polski}},
  \bibinfo {author} {\bibfnamefont {A.}~\bibnamefont {Thomson}}, \bibinfo
  {author} {\bibfnamefont {{\'E}.}~\bibnamefont {Lantagne-Hurtubise}}, \bibinfo
  {author} {\bibfnamefont {C.}~\bibnamefont {Lewandowski}}, \bibinfo {author}
  {\bibfnamefont {H.}~\bibnamefont {Zhou}}, \bibinfo {author} {\bibfnamefont
  {K.}~\bibnamefont {Watanabe}}, \bibinfo {author} {\bibfnamefont
  {T.}~\bibnamefont {Taniguchi}}, \bibinfo {author} {\bibfnamefont
  {J.}~\bibnamefont {Alicea}},\ and\ \bibinfo {author} {\bibfnamefont
  {S.}~\bibnamefont {Nadj-Perge}},\ }\bibfield  {title} {\bibinfo {title}
  {Enhanced superconductivity in spin--orbit proximitized bilayer graphene},\
  }\href@noop {} {\bibfield  {journal} {\bibinfo  {journal} {Nature}\ }\textbf
  {\bibinfo {volume} {613}},\ \bibinfo {pages} {268} (\bibinfo {year}
  {2023})}\BibitemShut {NoStop}%
\bibitem [{\citenamefont {Winterer}\ \emph {et~al.}(2023)\citenamefont
  {Winterer}, \citenamefont {Geisenhof}, \citenamefont {Fernandez},
  \citenamefont {Seiler}, \citenamefont {Zhang},\ and\ \citenamefont
  {Weitz}}]{winterer2023ferroelectric}%
  \BibitemOpen
  \bibfield  {author} {\bibinfo {author} {\bibfnamefont {F.}~\bibnamefont
  {Winterer}}, \bibinfo {author} {\bibfnamefont {F.~R.}\ \bibnamefont
  {Geisenhof}}, \bibinfo {author} {\bibfnamefont {N.}~\bibnamefont
  {Fernandez}}, \bibinfo {author} {\bibfnamefont {A.~M.}\ \bibnamefont
  {Seiler}}, \bibinfo {author} {\bibfnamefont {F.}~\bibnamefont {Zhang}},\ and\
  \bibinfo {author} {\bibfnamefont {R.~T.}\ \bibnamefont {Weitz}},\ }\href@noop
  {} {\bibinfo {title} {Ferroelectric and anomalous quantum hall states in bare
  rhombohedral trilayer graphene}} (\bibinfo {year} {2023}),\ \Eprint
  {https://arxiv.org/abs/2305.04950} {arXiv:2305.04950 [cond-mat.mes-hall]}
  \BibitemShut {NoStop}%
\bibitem [{\citenamefont {Seiler}\ \emph {et~al.}(2023)\citenamefont {Seiler},
  \citenamefont {Statz}, \citenamefont {Weimer}, \citenamefont {Jacobsen},
  \citenamefont {Watanabe}, \citenamefont {Taniguchi}, \citenamefont {Dong},
  \citenamefont {Levitov},\ and\ \citenamefont
  {Weitz}}]{seiler2023interactiondriven}%
  \BibitemOpen
  \bibfield  {author} {\bibinfo {author} {\bibfnamefont {A.~M.}\ \bibnamefont
  {Seiler}}, \bibinfo {author} {\bibfnamefont {M.}~\bibnamefont {Statz}},
  \bibinfo {author} {\bibfnamefont {I.}~\bibnamefont {Weimer}}, \bibinfo
  {author} {\bibfnamefont {N.}~\bibnamefont {Jacobsen}}, \bibinfo {author}
  {\bibfnamefont {K.}~\bibnamefont {Watanabe}}, \bibinfo {author}
  {\bibfnamefont {T.}~\bibnamefont {Taniguchi}}, \bibinfo {author}
  {\bibfnamefont {Z.}~\bibnamefont {Dong}}, \bibinfo {author} {\bibfnamefont
  {L.~S.}\ \bibnamefont {Levitov}},\ and\ \bibinfo {author} {\bibfnamefont
  {R.~T.}\ \bibnamefont {Weitz}},\ }\href@noop {} {\bibinfo {title}
  {Interaction-driven (quasi-) insulating ground states of gapped
  electron-doped bilayer graphene}} (\bibinfo {year} {2023}),\ \Eprint
  {https://arxiv.org/abs/2308.00827} {arXiv:2308.00827 [cond-mat.str-el]}
  \BibitemShut {NoStop}%
\bibitem [{\citenamefont {Seiler}\ \emph
  {et~al.}(2024{\natexlab{a}})\citenamefont {Seiler}, \citenamefont
  {Zhumagulov}, \citenamefont {Zollner}, \citenamefont {Yoon}, \citenamefont
  {Urbaniak}, \citenamefont {Geisenhof}, \citenamefont {Watanabe},
  \citenamefont {Taniguchi}, \citenamefont {Fabian}, \citenamefont {Zhang}
  \emph {et~al.}}]{seiler2024layer}%
  \BibitemOpen
  \bibfield  {author} {\bibinfo {author} {\bibfnamefont {A.~M.}\ \bibnamefont
  {Seiler}}, \bibinfo {author} {\bibfnamefont {Y.}~\bibnamefont {Zhumagulov}},
  \bibinfo {author} {\bibfnamefont {K.}~\bibnamefont {Zollner}}, \bibinfo
  {author} {\bibfnamefont {C.}~\bibnamefont {Yoon}}, \bibinfo {author}
  {\bibfnamefont {D.}~\bibnamefont {Urbaniak}}, \bibinfo {author}
  {\bibfnamefont {F.~R.}\ \bibnamefont {Geisenhof}}, \bibinfo {author}
  {\bibfnamefont {K.}~\bibnamefont {Watanabe}}, \bibinfo {author}
  {\bibfnamefont {T.}~\bibnamefont {Taniguchi}}, \bibinfo {author}
  {\bibfnamefont {J.}~\bibnamefont {Fabian}}, \bibinfo {author} {\bibfnamefont
  {F.}~\bibnamefont {Zhang}}, \emph {et~al.},\ }\bibfield  {title} {\bibinfo
  {title} {Layer-selective spin-orbit coupling and strong correlation in
  bilayer graphene},\ }\href@noop {} {\bibfield  {journal} {\bibinfo  {journal}
  {arXiv preprint arXiv:2403.17140}\ } (\bibinfo {year}
  {2024}{\natexlab{a}})}\BibitemShut {NoStop}%
\bibitem [{\citenamefont {Li}\ \emph {et~al.}(2024)\citenamefont {Li},
  \citenamefont {Xu}, \citenamefont {Li}, \citenamefont {Li}, \citenamefont
  {Li}, \citenamefont {Watanabe}, \citenamefont {Taniguchi}, \citenamefont
  {Tong}, \citenamefont {Shen}, \citenamefont {Lu}, \citenamefont {Jia},
  \citenamefont {Wu}, \citenamefont {Liu},\ and\ \citenamefont
  {Li}}]{li2024tunable}%
  \BibitemOpen
  \bibfield  {author} {\bibinfo {author} {\bibfnamefont {C.}~\bibnamefont
  {Li}}, \bibinfo {author} {\bibfnamefont {F.}~\bibnamefont {Xu}}, \bibinfo
  {author} {\bibfnamefont {B.}~\bibnamefont {Li}}, \bibinfo {author}
  {\bibfnamefont {J.}~\bibnamefont {Li}}, \bibinfo {author} {\bibfnamefont
  {G.}~\bibnamefont {Li}}, \bibinfo {author} {\bibfnamefont {K.}~\bibnamefont
  {Watanabe}}, \bibinfo {author} {\bibfnamefont {T.}~\bibnamefont {Taniguchi}},
  \bibinfo {author} {\bibfnamefont {B.}~\bibnamefont {Tong}}, \bibinfo {author}
  {\bibfnamefont {J.}~\bibnamefont {Shen}}, \bibinfo {author} {\bibfnamefont
  {L.}~\bibnamefont {Lu}}, \bibinfo {author} {\bibfnamefont {J.}~\bibnamefont
  {Jia}}, \bibinfo {author} {\bibfnamefont {F.}~\bibnamefont {Wu}}, \bibinfo
  {author} {\bibfnamefont {X.}~\bibnamefont {Liu}},\ and\ \bibinfo {author}
  {\bibfnamefont {T.}~\bibnamefont {Li}},\ }\href@noop {} {\bibinfo {title}
  {Tunable superconductivity in electron- and hole-doped bernal bilayer
  graphene}} (\bibinfo {year} {2024}),\ \Eprint
  {https://arxiv.org/abs/2405.04479} {arXiv:2405.04479 [cond-mat.supr-con]}
  \BibitemShut {NoStop}%
\bibitem [{\citenamefont {Tsui}\ \emph {et~al.}(2024)\citenamefont {Tsui},
  \citenamefont {He}, \citenamefont {Hu}, \citenamefont {Lake}, \citenamefont
  {Wang}, \citenamefont {Watanabe}, \citenamefont {Taniguchi}, \citenamefont
  {Zaletel},\ and\ \citenamefont {Yazdani}}]{tsui2024direct}%
  \BibitemOpen
  \bibfield  {author} {\bibinfo {author} {\bibfnamefont {Y.-C.}\ \bibnamefont
  {Tsui}}, \bibinfo {author} {\bibfnamefont {M.}~\bibnamefont {He}}, \bibinfo
  {author} {\bibfnamefont {Y.}~\bibnamefont {Hu}}, \bibinfo {author}
  {\bibfnamefont {E.}~\bibnamefont {Lake}}, \bibinfo {author} {\bibfnamefont
  {T.}~\bibnamefont {Wang}}, \bibinfo {author} {\bibfnamefont {K.}~\bibnamefont
  {Watanabe}}, \bibinfo {author} {\bibfnamefont {T.}~\bibnamefont {Taniguchi}},
  \bibinfo {author} {\bibfnamefont {M.~P.}\ \bibnamefont {Zaletel}},\ and\
  \bibinfo {author} {\bibfnamefont {A.}~\bibnamefont {Yazdani}},\ }\bibfield
  {title} {\bibinfo {title} {Direct observation of a magnetic-field-induced
  wigner crystal},\ }\href@noop {} {\bibfield  {journal} {\bibinfo  {journal}
  {Nature}\ }\textbf {\bibinfo {volume} {628}},\ \bibinfo {pages} {287}
  (\bibinfo {year} {2024})}\BibitemShut {NoStop}%
\bibitem [{\citenamefont {Xie}\ \emph {et~al.}(2024)\citenamefont {Xie},
  \citenamefont {Wolf}, \citenamefont {Xu}, \citenamefont {Cui}, \citenamefont
  {Xiong}, \citenamefont {Ou}, \citenamefont {Hays}, \citenamefont {Holleis},
  \citenamefont {Guo}, \citenamefont {Sheekey}, \citenamefont {Patterson},
  \citenamefont {Arp}, \citenamefont {Watanabe}, \citenamefont {Taniguchi},
  \citenamefont {Tongay}, \citenamefont {Young}, \citenamefont {MacDonald},\
  and\ \citenamefont {Jin}}]{xie2024optical}%
  \BibitemOpen
  \bibfield  {author} {\bibinfo {author} {\bibfnamefont {T.}~\bibnamefont
  {Xie}}, \bibinfo {author} {\bibfnamefont {T.~M.}\ \bibnamefont {Wolf}},
  \bibinfo {author} {\bibfnamefont {S.}~\bibnamefont {Xu}}, \bibinfo {author}
  {\bibfnamefont {Z.}~\bibnamefont {Cui}}, \bibinfo {author} {\bibfnamefont
  {R.}~\bibnamefont {Xiong}}, \bibinfo {author} {\bibfnamefont
  {Y.}~\bibnamefont {Ou}}, \bibinfo {author} {\bibfnamefont {P.}~\bibnamefont
  {Hays}}, \bibinfo {author} {\bibfnamefont {L.~F.}\ \bibnamefont {Holleis}},
  \bibinfo {author} {\bibfnamefont {Y.}~\bibnamefont {Guo}}, \bibinfo {author}
  {\bibfnamefont {O.~I.}\ \bibnamefont {Sheekey}}, \bibinfo {author}
  {\bibfnamefont {C.}~\bibnamefont {Patterson}}, \bibinfo {author}
  {\bibfnamefont {T.}~\bibnamefont {Arp}}, \bibinfo {author} {\bibfnamefont
  {K.}~\bibnamefont {Watanabe}}, \bibinfo {author} {\bibfnamefont
  {T.}~\bibnamefont {Taniguchi}}, \bibinfo {author} {\bibfnamefont {S.~A.}\
  \bibnamefont {Tongay}}, \bibinfo {author} {\bibfnamefont {A.~F.}\
  \bibnamefont {Young}}, \bibinfo {author} {\bibfnamefont {A.~H.}\ \bibnamefont
  {MacDonald}},\ and\ \bibinfo {author} {\bibfnamefont {C.}~\bibnamefont
  {Jin}},\ }\href@noop {} {\bibinfo {title} {Optical imaging of flavor order in
  flat band graphene}} (\bibinfo {year} {2024}),\ \Eprint
  {https://arxiv.org/abs/2405.08074} {arXiv:2405.08074 [cond-mat.mes-hall]}
  \BibitemShut {NoStop}%
\bibitem [{\citenamefont {Seiler}\ \emph
  {et~al.}(2024{\natexlab{b}})\citenamefont {Seiler}, \citenamefont
  {Zhumagulov}, \citenamefont {Zollner}, \citenamefont {Yoon}, \citenamefont
  {Urbaniak}, \citenamefont {Geisenhof}, \citenamefont {Watanabe},
  \citenamefont {Taniguchi}, \citenamefont {Fabian}, \citenamefont {Zhang},\
  and\ \citenamefont {Weitz}}]{seiler2024layerselective}%
  \BibitemOpen
  \bibfield  {author} {\bibinfo {author} {\bibfnamefont {A.~M.}\ \bibnamefont
  {Seiler}}, \bibinfo {author} {\bibfnamefont {Y.}~\bibnamefont {Zhumagulov}},
  \bibinfo {author} {\bibfnamefont {K.}~\bibnamefont {Zollner}}, \bibinfo
  {author} {\bibfnamefont {C.}~\bibnamefont {Yoon}}, \bibinfo {author}
  {\bibfnamefont {D.}~\bibnamefont {Urbaniak}}, \bibinfo {author}
  {\bibfnamefont {F.~R.}\ \bibnamefont {Geisenhof}}, \bibinfo {author}
  {\bibfnamefont {K.}~\bibnamefont {Watanabe}}, \bibinfo {author}
  {\bibfnamefont {T.}~\bibnamefont {Taniguchi}}, \bibinfo {author}
  {\bibfnamefont {J.}~\bibnamefont {Fabian}}, \bibinfo {author} {\bibfnamefont
  {F.}~\bibnamefont {Zhang}},\ and\ \bibinfo {author} {\bibfnamefont {R.~T.}\
  \bibnamefont {Weitz}},\ }\href@noop {} {\bibinfo {title} {Layer-selective
  spin-orbit coupling and strong correlation in bilayer graphene}} (\bibinfo
  {year} {2024}{\natexlab{b}}),\ \Eprint {https://arxiv.org/abs/2403.17140}
  {arXiv:2403.17140 [cond-mat.mes-hall]} \BibitemShut {NoStop}%
\bibitem [{\citenamefont {Chatterjee}\ \emph {et~al.}(2022)\citenamefont
  {Chatterjee}, \citenamefont {Wang}, \citenamefont {Berg},\ and\ \citenamefont
  {Zaletel}}]{chatterjee2022inter}%
  \BibitemOpen
  \bibfield  {author} {\bibinfo {author} {\bibfnamefont {S.}~\bibnamefont
  {Chatterjee}}, \bibinfo {author} {\bibfnamefont {T.}~\bibnamefont {Wang}},
  \bibinfo {author} {\bibfnamefont {E.}~\bibnamefont {Berg}},\ and\ \bibinfo
  {author} {\bibfnamefont {M.~P.}\ \bibnamefont {Zaletel}},\ }\bibfield
  {title} {\bibinfo {title} {Inter-valley coherent order and isospin
  fluctuation mediated superconductivity in rhombohedral trilayer graphene},\
  }\href@noop {} {\bibfield  {journal} {\bibinfo  {journal} {Nature
  Communications}\ }\textbf {\bibinfo {volume} {13}},\ \bibinfo {pages} {6013}
  (\bibinfo {year} {2022})}\BibitemShut {NoStop}%
\bibitem [{\citenamefont {Szab{\'o}}\ and\ \citenamefont
  {Roy}(2022{\natexlab{a}})}]{szabo2022competing}%
  \BibitemOpen
  \bibfield  {author} {\bibinfo {author} {\bibfnamefont {A.~L.}\ \bibnamefont
  {Szab{\'o}}}\ and\ \bibinfo {author} {\bibfnamefont {B.}~\bibnamefont
  {Roy}},\ }\bibfield  {title} {\bibinfo {title} {Competing orders and cascade
  of degeneracy lifting in doped bernal bilayer graphene},\ }\href@noop {}
  {\bibfield  {journal} {\bibinfo  {journal} {Physical Review B}\ }\textbf
  {\bibinfo {volume} {105}},\ \bibinfo {pages} {L201107} (\bibinfo {year}
  {2022}{\natexlab{a}})}\BibitemShut {NoStop}%
\bibitem [{\citenamefont {Szab{\'o}}\ and\ \citenamefont
  {Roy}(2022{\natexlab{b}})}]{szabo2022metals}%
  \BibitemOpen
  \bibfield  {author} {\bibinfo {author} {\bibfnamefont {A.~L.}\ \bibnamefont
  {Szab{\'o}}}\ and\ \bibinfo {author} {\bibfnamefont {B.}~\bibnamefont
  {Roy}},\ }\bibfield  {title} {\bibinfo {title} {Metals, fractional metals,
  and superconductivity in rhombohedral trilayer graphene},\ }\href@noop {}
  {\bibfield  {journal} {\bibinfo  {journal} {Physical Review B}\ }\textbf
  {\bibinfo {volume} {105}},\ \bibinfo {pages} {L081407} (\bibinfo {year}
  {2022}{\natexlab{b}})}\BibitemShut {NoStop}%
\bibitem [{\citenamefont {You}\ and\ \citenamefont
  {Vishwanath}(2022)}]{PhysRevB.105.134524}%
  \BibitemOpen
  \bibfield  {author} {\bibinfo {author} {\bibfnamefont {Y.-Z.}\ \bibnamefont
  {You}}\ and\ \bibinfo {author} {\bibfnamefont {A.}~\bibnamefont
  {Vishwanath}},\ }\bibfield  {title} {\bibinfo {title} {Kohn-luttinger
  superconductivity and intervalley coherence in rhombohedral trilayer
  graphene},\ }\href {https://doi.org/10.1103/PhysRevB.105.134524} {\bibfield
  {journal} {\bibinfo  {journal} {Phys. Rev. B}\ }\textbf {\bibinfo {volume}
  {105}},\ \bibinfo {pages} {134524} (\bibinfo {year} {2022})}\BibitemShut
  {NoStop}%
\bibitem [{\citenamefont {Li}\ \emph {et~al.}(2023)\citenamefont {Li},
  \citenamefont {Kuang}, \citenamefont {Jimeno-Pozo}, \citenamefont
  {Sainz-Cruz}, \citenamefont {Zhan}, \citenamefont {Yuan},\ and\ \citenamefont
  {Guinea}}]{li2023charge}%
  \BibitemOpen
  \bibfield  {author} {\bibinfo {author} {\bibfnamefont {Z.}~\bibnamefont
  {Li}}, \bibinfo {author} {\bibfnamefont {X.}~\bibnamefont {Kuang}}, \bibinfo
  {author} {\bibfnamefont {A.}~\bibnamefont {Jimeno-Pozo}}, \bibinfo {author}
  {\bibfnamefont {H.}~\bibnamefont {Sainz-Cruz}}, \bibinfo {author}
  {\bibfnamefont {Z.}~\bibnamefont {Zhan}}, \bibinfo {author} {\bibfnamefont
  {S.}~\bibnamefont {Yuan}},\ and\ \bibinfo {author} {\bibfnamefont
  {F.}~\bibnamefont {Guinea}},\ }\bibfield  {title} {\bibinfo {title} {Charge
  fluctuations, phonons and superconductivity in multilayer graphene},\
  }\href@noop {} {\bibfield  {journal} {\bibinfo  {journal} {arXiv preprint
  arXiv:2303.17286}\ } (\bibinfo {year} {2023})}\BibitemShut {NoStop}%
\bibitem [{\citenamefont {Pantaleon}\ \emph {et~al.}(2022)\citenamefont
  {Pantaleon}, \citenamefont {Jimeno-Pozo}, \citenamefont {Sainz-Cruz},
  \citenamefont {Cea}, \citenamefont {Phong},\ and\ \citenamefont
  {Guinea}}]{pantaleon2022superconductivity}%
  \BibitemOpen
  \bibfield  {author} {\bibinfo {author} {\bibfnamefont {P.~A.}\ \bibnamefont
  {Pantaleon}}, \bibinfo {author} {\bibfnamefont {A.}~\bibnamefont
  {Jimeno-Pozo}}, \bibinfo {author} {\bibfnamefont {H.}~\bibnamefont
  {Sainz-Cruz}}, \bibinfo {author} {\bibfnamefont {T.}~\bibnamefont {Cea}},
  \bibinfo {author} {\bibfnamefont {V.~T.}\ \bibnamefont {Phong}},\ and\
  \bibinfo {author} {\bibfnamefont {F.}~\bibnamefont {Guinea}},\ }\bibfield
  {title} {\bibinfo {title} {Superconductivity and correlated phases in
  bilayer, trilayer graphene and related structures},\ }\href@noop {}
  {\bibfield  {journal} {\bibinfo  {journal} {arXiv preprint arXiv:2211.02880}\
  } (\bibinfo {year} {2022})}\BibitemShut {NoStop}%
\bibitem [{\citenamefont {Pantale{\'o}n}\ \emph {et~al.}(2023)\citenamefont
  {Pantale{\'o}n}, \citenamefont {Jimeno-Pozo}, \citenamefont {Sainz-Cruz},
  \citenamefont {Phong}, \citenamefont {Cea},\ and\ \citenamefont
  {Guinea}}]{pantaleon2023superconductivity}%
  \BibitemOpen
  \bibfield  {author} {\bibinfo {author} {\bibfnamefont {P.~A.}\ \bibnamefont
  {Pantale{\'o}n}}, \bibinfo {author} {\bibfnamefont {A.}~\bibnamefont
  {Jimeno-Pozo}}, \bibinfo {author} {\bibfnamefont {H.}~\bibnamefont
  {Sainz-Cruz}}, \bibinfo {author} {\bibfnamefont {V.~T.}\ \bibnamefont
  {Phong}}, \bibinfo {author} {\bibfnamefont {T.}~\bibnamefont {Cea}},\ and\
  \bibinfo {author} {\bibfnamefont {F.}~\bibnamefont {Guinea}},\ }\bibfield
  {title} {\bibinfo {title} {Superconductivity and correlated phases in
  non-twisted bilayer and trilayer graphene},\ }\href@noop {} {\bibfield
  {journal} {\bibinfo  {journal} {Nature Reviews Physics}\ ,\ \bibinfo {pages}
  {1}} (\bibinfo {year} {2023})}\BibitemShut {NoStop}%
\bibitem [{\citenamefont {Cea}\ \emph {et~al.}(2022)\citenamefont {Cea},
  \citenamefont {Pantale{\'o}n}, \citenamefont {Phong},\ and\ \citenamefont
  {Guinea}}]{cea2022superconductivity}%
  \BibitemOpen
  \bibfield  {author} {\bibinfo {author} {\bibfnamefont {T.}~\bibnamefont
  {Cea}}, \bibinfo {author} {\bibfnamefont {P.~A.}\ \bibnamefont
  {Pantale{\'o}n}}, \bibinfo {author} {\bibfnamefont {V.~T.}\ \bibnamefont
  {Phong}},\ and\ \bibinfo {author} {\bibfnamefont {F.}~\bibnamefont
  {Guinea}},\ }\bibfield  {title} {\bibinfo {title} {Superconductivity from
  repulsive interactions in rhombohedral trilayer graphene: A
  kohn-luttinger-like mechanism},\ }\href@noop {} {\bibfield  {journal}
  {\bibinfo  {journal} {Physical Review B}\ }\textbf {\bibinfo {volume}
  {105}},\ \bibinfo {pages} {075432} (\bibinfo {year} {2022})}\BibitemShut
  {NoStop}%
\bibitem [{\citenamefont {Jimeno-Pozo}\ \emph {et~al.}(2023)\citenamefont
  {Jimeno-Pozo}, \citenamefont {Sainz-Cruz}, \citenamefont {Cea}, \citenamefont
  {Pantale{\'o}n},\ and\ \citenamefont {Guinea}}]{jimeno2023superconductivity}%
  \BibitemOpen
  \bibfield  {author} {\bibinfo {author} {\bibfnamefont {A.}~\bibnamefont
  {Jimeno-Pozo}}, \bibinfo {author} {\bibfnamefont {H.}~\bibnamefont
  {Sainz-Cruz}}, \bibinfo {author} {\bibfnamefont {T.}~\bibnamefont {Cea}},
  \bibinfo {author} {\bibfnamefont {P.~A.}\ \bibnamefont {Pantale{\'o}n}},\
  and\ \bibinfo {author} {\bibfnamefont {F.}~\bibnamefont {Guinea}},\
  }\bibfield  {title} {\bibinfo {title} {Superconductivity from electronic
  interactions and spin-orbit enhancement in bilayer and trilayer graphene},\
  }\href@noop {} {\bibfield  {journal} {\bibinfo  {journal} {Physical Review
  B}\ }\textbf {\bibinfo {volume} {107}},\ \bibinfo {pages} {L161106} (\bibinfo
  {year} {2023})}\BibitemShut {NoStop}%
\bibitem [{\citenamefont {Cea}(2023)}]{PhysRevB.107.L041111}%
  \BibitemOpen
  \bibfield  {author} {\bibinfo {author} {\bibfnamefont {T.}~\bibnamefont
  {Cea}},\ }\bibfield  {title} {\bibinfo {title} {Superconductivity induced by
  the intervalley coulomb scattering in a few layers of graphene},\ }\href
  {https://doi.org/10.1103/PhysRevB.107.L041111} {\bibfield  {journal}
  {\bibinfo  {journal} {Phys. Rev. B}\ }\textbf {\bibinfo {volume} {107}},\
  \bibinfo {pages} {L041111} (\bibinfo {year} {2023})}\BibitemShut {NoStop}%
\bibitem [{\citenamefont {Dai}\ \emph {et~al.}(2022)\citenamefont {Dai},
  \citenamefont {Ma}, \citenamefont {Zhang},\ and\ \citenamefont
  {Ma}}]{dai2022quantum}%
  \BibitemOpen
  \bibfield  {author} {\bibinfo {author} {\bibfnamefont {H.}~\bibnamefont
  {Dai}}, \bibinfo {author} {\bibfnamefont {R.}~\bibnamefont {Ma}}, \bibinfo
  {author} {\bibfnamefont {X.}~\bibnamefont {Zhang}},\ and\ \bibinfo {author}
  {\bibfnamefont {T.}~\bibnamefont {Ma}},\ }\href@noop {} {\bibinfo {title}
  {Quantum monte carlo study of superconductivity in rhombohedral trilayer
  graphene under an electric field}} (\bibinfo {year} {2022}),\ \Eprint
  {https://arxiv.org/abs/2204.06222} {arXiv:2204.06222 [cond-mat.str-el]}
  \BibitemShut {NoStop}%
\bibitem [{\citenamefont {Ghazaryan}\ \emph {et~al.}(2023)\citenamefont
  {Ghazaryan}, \citenamefont {Holder}, \citenamefont {Berg},\ and\
  \citenamefont {Serbyn}}]{ghazaryan2023multilayer}%
  \BibitemOpen
  \bibfield  {author} {\bibinfo {author} {\bibfnamefont {A.}~\bibnamefont
  {Ghazaryan}}, \bibinfo {author} {\bibfnamefont {T.}~\bibnamefont {Holder}},
  \bibinfo {author} {\bibfnamefont {E.}~\bibnamefont {Berg}},\ and\ \bibinfo
  {author} {\bibfnamefont {M.}~\bibnamefont {Serbyn}},\ }\bibfield  {title}
  {\bibinfo {title} {Multilayer graphenes as a platform for interaction-driven
  physics and topological superconductivity},\ }\href@noop {} {\bibfield
  {journal} {\bibinfo  {journal} {Physical Review B}\ }\textbf {\bibinfo
  {volume} {107}},\ \bibinfo {pages} {104502} (\bibinfo {year}
  {2023})}\BibitemShut {NoStop}%
\bibitem [{\citenamefont {Qin}\ \emph {et~al.}(2023)\citenamefont {Qin},
  \citenamefont {Huang}, \citenamefont {Wolf}, \citenamefont {Wei},
  \citenamefont {Blinov},\ and\ \citenamefont {MacDonald}}]{wei2023}%
  \BibitemOpen
  \bibfield  {author} {\bibinfo {author} {\bibfnamefont {W.}~\bibnamefont
  {Qin}}, \bibinfo {author} {\bibfnamefont {C.}~\bibnamefont {Huang}}, \bibinfo
  {author} {\bibfnamefont {T.}~\bibnamefont {Wolf}}, \bibinfo {author}
  {\bibfnamefont {N.}~\bibnamefont {Wei}}, \bibinfo {author} {\bibfnamefont
  {I.}~\bibnamefont {Blinov}},\ and\ \bibinfo {author} {\bibfnamefont {A.~H.}\
  \bibnamefont {MacDonald}},\ }\bibfield  {title} {\bibinfo {title} {Functional
  renormalization group study of superconductivity in rhombohedral trilayer
  graphene},\ }\href {https://doi.org/10.1103/PhysRevLett.130.146001}
  {\bibfield  {journal} {\bibinfo  {journal} {Phys. Rev. Lett.}\ }\textbf
  {\bibinfo {volume} {130}},\ \bibinfo {pages} {146001} (\bibinfo {year}
  {2023})}\BibitemShut {NoStop}%
\bibitem [{\citenamefont {Wagner}\ \emph {et~al.}(2023)\citenamefont {Wagner},
  \citenamefont {Kwan}, \citenamefont {Bultinck}, \citenamefont {Simon},\ and\
  \citenamefont {Parameswaran}}]{wagner2023superconductivity}%
  \BibitemOpen
  \bibfield  {author} {\bibinfo {author} {\bibfnamefont {G.}~\bibnamefont
  {Wagner}}, \bibinfo {author} {\bibfnamefont {Y.~H.}\ \bibnamefont {Kwan}},
  \bibinfo {author} {\bibfnamefont {N.}~\bibnamefont {Bultinck}}, \bibinfo
  {author} {\bibfnamefont {S.~H.}\ \bibnamefont {Simon}},\ and\ \bibinfo
  {author} {\bibfnamefont {S.}~\bibnamefont {Parameswaran}},\ }\bibfield
  {title} {\bibinfo {title} {Superconductivity from repulsive interactions in
  bernal-stacked bilayer graphene},\ }\href@noop {} {\bibfield  {journal}
  {\bibinfo  {journal} {arXiv preprint arXiv:2302.00682}\ } (\bibinfo {year}
  {2023})}\BibitemShut {NoStop}%
\bibitem [{\citenamefont {Fischer}\ \emph {et~al.}(2024)\citenamefont
  {Fischer}, \citenamefont {Klebl}, \citenamefont {Profe}, \citenamefont
  {Rothstein}, \citenamefont {Waldecker}, \citenamefont {Beschoten},
  \citenamefont {Wehling},\ and\ \citenamefont {Kennes}}]{fischer2023spin}%
  \BibitemOpen
  \bibfield  {author} {\bibinfo {author} {\bibfnamefont {A.}~\bibnamefont
  {Fischer}}, \bibinfo {author} {\bibfnamefont {L.}~\bibnamefont {Klebl}},
  \bibinfo {author} {\bibfnamefont {J.~B.}\ \bibnamefont {Profe}}, \bibinfo
  {author} {\bibfnamefont {A.}~\bibnamefont {Rothstein}}, \bibinfo {author}
  {\bibfnamefont {L.}~\bibnamefont {Waldecker}}, \bibinfo {author}
  {\bibfnamefont {B.}~\bibnamefont {Beschoten}}, \bibinfo {author}
  {\bibfnamefont {T.~O.}\ \bibnamefont {Wehling}},\ and\ \bibinfo {author}
  {\bibfnamefont {D.~M.}\ \bibnamefont {Kennes}},\ }\bibfield  {title}
  {\bibinfo {title} {Spin and charge fluctuation induced pairing in abcb
  tetralayer graphene},\ }\href@noop {} {\bibfield  {journal} {\bibinfo
  {journal} {Physical Review Research}\ }\textbf {\bibinfo {volume} {6}},\
  \bibinfo {pages} {L012003} (\bibinfo {year} {2024})}\BibitemShut {NoStop}%
\bibitem [{\citenamefont {Zhumagulov}\ \emph {et~al.}(2023)\citenamefont
  {Zhumagulov}, \citenamefont {Kochan},\ and\ \citenamefont
  {Fabian}}]{zhumagulov2023emergent}%
  \BibitemOpen
  \bibfield  {author} {\bibinfo {author} {\bibfnamefont {Y.}~\bibnamefont
  {Zhumagulov}}, \bibinfo {author} {\bibfnamefont {D.}~\bibnamefont {Kochan}},\
  and\ \bibinfo {author} {\bibfnamefont {J.}~\bibnamefont {Fabian}},\
  }\href@noop {} {\bibinfo {title} {Emergent correlated phases in rhombohedral
  trilayer graphene induced by proximity spin-orbit and exchange coupling}}
  (\bibinfo {year} {2023}),\ \Eprint {https://arxiv.org/abs/2305.14277}
  {arXiv:2305.14277 [cond-mat.str-el]} \BibitemShut {NoStop}%
\bibitem [{\citenamefont {Zhou}\ \emph
  {et~al.}(2021{\natexlab{a}})\citenamefont {Zhou}, \citenamefont {Xie},
  \citenamefont {Ghazaryan}, \citenamefont {Holder}, \citenamefont {Ehrets},
  \citenamefont {Spanton}, \citenamefont {Taniguchi}, \citenamefont {Watanabe},
  \citenamefont {Berg}, \citenamefont {Serbyn} \emph {et~al.}}]{zhou2021half}%
  \BibitemOpen
  \bibfield  {author} {\bibinfo {author} {\bibfnamefont {H.}~\bibnamefont
  {Zhou}}, \bibinfo {author} {\bibfnamefont {T.}~\bibnamefont {Xie}}, \bibinfo
  {author} {\bibfnamefont {A.}~\bibnamefont {Ghazaryan}}, \bibinfo {author}
  {\bibfnamefont {T.}~\bibnamefont {Holder}}, \bibinfo {author} {\bibfnamefont
  {J.~R.}\ \bibnamefont {Ehrets}}, \bibinfo {author} {\bibfnamefont {E.~M.}\
  \bibnamefont {Spanton}}, \bibinfo {author} {\bibfnamefont {T.}~\bibnamefont
  {Taniguchi}}, \bibinfo {author} {\bibfnamefont {K.}~\bibnamefont {Watanabe}},
  \bibinfo {author} {\bibfnamefont {E.}~\bibnamefont {Berg}}, \bibinfo {author}
  {\bibfnamefont {M.}~\bibnamefont {Serbyn}}, \emph {et~al.},\ }\bibfield
  {title} {\bibinfo {title} {Half-and quarter-metals in rhombohedral trilayer
  graphene},\ }\href@noop {} {\bibfield  {journal} {\bibinfo  {journal}
  {Nature}\ }\textbf {\bibinfo {volume} {598}},\ \bibinfo {pages} {429}
  (\bibinfo {year} {2021}{\natexlab{a}})}\BibitemShut {NoStop}%
\bibitem [{\citenamefont {Zhou}\ \emph
  {et~al.}(2021{\natexlab{b}})\citenamefont {Zhou}, \citenamefont {Xie},
  \citenamefont {Taniguchi}, \citenamefont {Watanabe},\ and\ \citenamefont
  {Young}}]{zhou2021superconductivity}%
  \BibitemOpen
  \bibfield  {author} {\bibinfo {author} {\bibfnamefont {H.}~\bibnamefont
  {Zhou}}, \bibinfo {author} {\bibfnamefont {T.}~\bibnamefont {Xie}}, \bibinfo
  {author} {\bibfnamefont {T.}~\bibnamefont {Taniguchi}}, \bibinfo {author}
  {\bibfnamefont {K.}~\bibnamefont {Watanabe}},\ and\ \bibinfo {author}
  {\bibfnamefont {A.~F.}\ \bibnamefont {Young}},\ }\bibfield  {title} {\bibinfo
  {title} {Superconductivity in rhombohedral trilayer graphene},\ }\href@noop
  {} {\bibfield  {journal} {\bibinfo  {journal} {Nature}\ }\textbf {\bibinfo
  {volume} {598}},\ \bibinfo {pages} {434} (\bibinfo {year}
  {2021}{\natexlab{b}})}\BibitemShut {NoStop}%
\bibitem [{\citenamefont {Cao}\ \emph {et~al.}(2018)\citenamefont {Cao},
  \citenamefont {Fatemi}, \citenamefont {Fang}, \citenamefont {Watanabe},
  \citenamefont {Taniguchi}, \citenamefont {Kaxiras},\ and\ \citenamefont
  {Jarillo-Herrero}}]{cao_unconventional_2018}%
  \BibitemOpen
  \bibfield  {author} {\bibinfo {author} {\bibfnamefont {Y.}~\bibnamefont
  {Cao}}, \bibinfo {author} {\bibfnamefont {V.}~\bibnamefont {Fatemi}},
  \bibinfo {author} {\bibfnamefont {S.}~\bibnamefont {Fang}}, \bibinfo {author}
  {\bibfnamefont {K.}~\bibnamefont {Watanabe}}, \bibinfo {author}
  {\bibfnamefont {T.}~\bibnamefont {Taniguchi}}, \bibinfo {author}
  {\bibfnamefont {E.}~\bibnamefont {Kaxiras}},\ and\ \bibinfo {author}
  {\bibfnamefont {P.}~\bibnamefont {Jarillo-Herrero}},\ }\bibfield  {title}
  {\bibinfo {title} {Unconventional superconductivity in magic-angle graphene
  superlattices},\ }\href {https://doi.org/10.1038/nature26160} {\bibfield
  {journal} {\bibinfo  {journal} {Nature}\ }\textbf {\bibinfo {volume} {556}},\
  \bibinfo {pages} {43} (\bibinfo {year} {2018})}\BibitemShut {NoStop}%
\bibitem [{\citenamefont {Yankowitz}\ \emph {et~al.}(2019)\citenamefont
  {Yankowitz}, \citenamefont {Chen}, \citenamefont {Polshyn}, \citenamefont
  {Zhang}, \citenamefont {Watanabe}, \citenamefont {Taniguchi}, \citenamefont
  {Graf}, \citenamefont {Young},\ and\ \citenamefont
  {Dean}}]{yankowitz2019tuning}%
  \BibitemOpen
  \bibfield  {author} {\bibinfo {author} {\bibfnamefont {M.}~\bibnamefont
  {Yankowitz}}, \bibinfo {author} {\bibfnamefont {S.}~\bibnamefont {Chen}},
  \bibinfo {author} {\bibfnamefont {H.}~\bibnamefont {Polshyn}}, \bibinfo
  {author} {\bibfnamefont {Y.}~\bibnamefont {Zhang}}, \bibinfo {author}
  {\bibfnamefont {K.}~\bibnamefont {Watanabe}}, \bibinfo {author}
  {\bibfnamefont {T.}~\bibnamefont {Taniguchi}}, \bibinfo {author}
  {\bibfnamefont {D.}~\bibnamefont {Graf}}, \bibinfo {author} {\bibfnamefont
  {A.~F.}\ \bibnamefont {Young}},\ and\ \bibinfo {author} {\bibfnamefont
  {C.~R.}\ \bibnamefont {Dean}},\ }\bibfield  {title} {\bibinfo {title} {Tuning
  superconductivity in twisted bilayer graphene},\ }\href@noop {} {\bibfield
  {journal} {\bibinfo  {journal} {Science}\ }\textbf {\bibinfo {volume}
  {363}},\ \bibinfo {pages} {1059} (\bibinfo {year} {2019})}\BibitemShut
  {NoStop}%
\bibitem [{\citenamefont {Lu}\ \emph {et~al.}(2019)\citenamefont {Lu},
  \citenamefont {Stepanov}, \citenamefont {Yang}, \citenamefont {Xie},
  \citenamefont {Aamir}, \citenamefont {Das}, \citenamefont {Urgell},
  \citenamefont {Watanabe}, \citenamefont {Taniguchi}, \citenamefont {Zhang}
  \emph {et~al.}}]{lu2019superconductors}%
  \BibitemOpen
  \bibfield  {author} {\bibinfo {author} {\bibfnamefont {X.}~\bibnamefont
  {Lu}}, \bibinfo {author} {\bibfnamefont {P.}~\bibnamefont {Stepanov}},
  \bibinfo {author} {\bibfnamefont {W.}~\bibnamefont {Yang}}, \bibinfo {author}
  {\bibfnamefont {M.}~\bibnamefont {Xie}}, \bibinfo {author} {\bibfnamefont
  {M.~A.}\ \bibnamefont {Aamir}}, \bibinfo {author} {\bibfnamefont
  {I.}~\bibnamefont {Das}}, \bibinfo {author} {\bibfnamefont {C.}~\bibnamefont
  {Urgell}}, \bibinfo {author} {\bibfnamefont {K.}~\bibnamefont {Watanabe}},
  \bibinfo {author} {\bibfnamefont {T.}~\bibnamefont {Taniguchi}}, \bibinfo
  {author} {\bibfnamefont {G.}~\bibnamefont {Zhang}}, \emph {et~al.},\
  }\bibfield  {title} {\bibinfo {title} {Superconductors, orbital magnets and
  correlated states in magic-angle bilayer graphene},\ }\href@noop {}
  {\bibfield  {journal} {\bibinfo  {journal} {Nature}\ }\textbf {\bibinfo
  {volume} {574}},\ \bibinfo {pages} {653} (\bibinfo {year}
  {2019})}\BibitemShut {NoStop}%
\bibitem [{\citenamefont {Saito}\ \emph {et~al.}(2020)\citenamefont {Saito},
  \citenamefont {Ge}, \citenamefont {Watanabe}, \citenamefont {Taniguchi},\
  and\ \citenamefont {Young}}]{saito2020independent}%
  \BibitemOpen
  \bibfield  {author} {\bibinfo {author} {\bibfnamefont {Y.}~\bibnamefont
  {Saito}}, \bibinfo {author} {\bibfnamefont {J.}~\bibnamefont {Ge}}, \bibinfo
  {author} {\bibfnamefont {K.}~\bibnamefont {Watanabe}}, \bibinfo {author}
  {\bibfnamefont {T.}~\bibnamefont {Taniguchi}},\ and\ \bibinfo {author}
  {\bibfnamefont {A.~F.}\ \bibnamefont {Young}},\ }\bibfield  {title} {\bibinfo
  {title} {Independent superconductors and correlated insulators in twisted
  bilayer graphene},\ }\href@noop {} {\bibfield  {journal} {\bibinfo  {journal}
  {Nature Physics}\ }\textbf {\bibinfo {volume} {16}},\ \bibinfo {pages} {926}
  (\bibinfo {year} {2020})}\BibitemShut {NoStop}%
\bibitem [{\citenamefont {Stepanov}\ \emph {et~al.}(2020)\citenamefont
  {Stepanov}, \citenamefont {Das}, \citenamefont {Lu}, \citenamefont
  {Fahimniya}, \citenamefont {Watanabe}, \citenamefont {Taniguchi},
  \citenamefont {Koppens}, \citenamefont {Lischner}, \citenamefont {Levitov},\
  and\ \citenamefont {Efetov}}]{stepanov2020untying}%
  \BibitemOpen
  \bibfield  {author} {\bibinfo {author} {\bibfnamefont {P.}~\bibnamefont
  {Stepanov}}, \bibinfo {author} {\bibfnamefont {I.}~\bibnamefont {Das}},
  \bibinfo {author} {\bibfnamefont {X.}~\bibnamefont {Lu}}, \bibinfo {author}
  {\bibfnamefont {A.}~\bibnamefont {Fahimniya}}, \bibinfo {author}
  {\bibfnamefont {K.}~\bibnamefont {Watanabe}}, \bibinfo {author}
  {\bibfnamefont {T.}~\bibnamefont {Taniguchi}}, \bibinfo {author}
  {\bibfnamefont {F.~H.}\ \bibnamefont {Koppens}}, \bibinfo {author}
  {\bibfnamefont {J.}~\bibnamefont {Lischner}}, \bibinfo {author}
  {\bibfnamefont {L.}~\bibnamefont {Levitov}},\ and\ \bibinfo {author}
  {\bibfnamefont {D.~K.}\ \bibnamefont {Efetov}},\ }\bibfield  {title}
  {\bibinfo {title} {Untying the insulating and superconducting orders in
  magic-angle graphene},\ }\href@noop {} {\bibfield  {journal} {\bibinfo
  {journal} {Nature}\ }\textbf {\bibinfo {volume} {583}},\ \bibinfo {pages}
  {375} (\bibinfo {year} {2020})}\BibitemShut {NoStop}%
\bibitem [{\citenamefont {Oh}\ \emph {et~al.}(2021)\citenamefont {Oh},
  \citenamefont {Nuckolls}, \citenamefont {Wong}, \citenamefont {Lee},
  \citenamefont {Liu}, \citenamefont {Watanabe}, \citenamefont {Taniguchi},\
  and\ \citenamefont {Yazdani}}]{oh2021evidence}%
  \BibitemOpen
  \bibfield  {author} {\bibinfo {author} {\bibfnamefont {M.}~\bibnamefont
  {Oh}}, \bibinfo {author} {\bibfnamefont {K.~P.}\ \bibnamefont {Nuckolls}},
  \bibinfo {author} {\bibfnamefont {D.}~\bibnamefont {Wong}}, \bibinfo {author}
  {\bibfnamefont {R.~L.}\ \bibnamefont {Lee}}, \bibinfo {author} {\bibfnamefont
  {X.}~\bibnamefont {Liu}}, \bibinfo {author} {\bibfnamefont {K.}~\bibnamefont
  {Watanabe}}, \bibinfo {author} {\bibfnamefont {T.}~\bibnamefont
  {Taniguchi}},\ and\ \bibinfo {author} {\bibfnamefont {A.}~\bibnamefont
  {Yazdani}},\ }\bibfield  {title} {\bibinfo {title} {Evidence for
  unconventional superconductivity in twisted bilayer graphene},\ }\href@noop
  {} {\bibfield  {journal} {\bibinfo  {journal} {Nature}\ }\textbf {\bibinfo
  {volume} {600}},\ \bibinfo {pages} {240} (\bibinfo {year}
  {2021})}\BibitemShut {NoStop}%
\bibitem [{\citenamefont {Cao}\ \emph {et~al.}(2021{\natexlab{a}})\citenamefont
  {Cao}, \citenamefont {Rodan-Legrain}, \citenamefont {Park}, \citenamefont
  {Yuan}, \citenamefont {Watanabe}, \citenamefont {Taniguchi}, \citenamefont
  {Fernandes}, \citenamefont {Fu},\ and\ \citenamefont
  {Jarillo-Herrero}}]{cao2021nematicity}%
  \BibitemOpen
  \bibfield  {author} {\bibinfo {author} {\bibfnamefont {Y.}~\bibnamefont
  {Cao}}, \bibinfo {author} {\bibfnamefont {D.}~\bibnamefont {Rodan-Legrain}},
  \bibinfo {author} {\bibfnamefont {J.~M.}\ \bibnamefont {Park}}, \bibinfo
  {author} {\bibfnamefont {N.~F.}\ \bibnamefont {Yuan}}, \bibinfo {author}
  {\bibfnamefont {K.}~\bibnamefont {Watanabe}}, \bibinfo {author}
  {\bibfnamefont {T.}~\bibnamefont {Taniguchi}}, \bibinfo {author}
  {\bibfnamefont {R.~M.}\ \bibnamefont {Fernandes}}, \bibinfo {author}
  {\bibfnamefont {L.}~\bibnamefont {Fu}},\ and\ \bibinfo {author}
  {\bibfnamefont {P.}~\bibnamefont {Jarillo-Herrero}},\ }\bibfield  {title}
  {\bibinfo {title} {Nematicity and competing orders in superconducting
  magic-angle graphene},\ }\href@noop {} {\bibfield  {journal} {\bibinfo
  {journal} {science}\ }\textbf {\bibinfo {volume} {372}},\ \bibinfo {pages}
  {264} (\bibinfo {year} {2021}{\natexlab{a}})}\BibitemShut {NoStop}%
\bibitem [{\citenamefont {Zhang}\ \emph {et~al.}(2021)\citenamefont {Zhang},
  \citenamefont {Polski}, \citenamefont {Lewandowski}, \citenamefont {Thomson},
  \citenamefont {Peng}, \citenamefont {Choi}, \citenamefont {Kim},
  \citenamefont {Watanabe}, \citenamefont {Taniguchi}, \citenamefont {Alicea}
  \emph {et~al.}}]{zhang2021ascendance}%
  \BibitemOpen
  \bibfield  {author} {\bibinfo {author} {\bibfnamefont {Y.}~\bibnamefont
  {Zhang}}, \bibinfo {author} {\bibfnamefont {R.}~\bibnamefont {Polski}},
  \bibinfo {author} {\bibfnamefont {C.}~\bibnamefont {Lewandowski}}, \bibinfo
  {author} {\bibfnamefont {A.}~\bibnamefont {Thomson}}, \bibinfo {author}
  {\bibfnamefont {Y.}~\bibnamefont {Peng}}, \bibinfo {author} {\bibfnamefont
  {Y.}~\bibnamefont {Choi}}, \bibinfo {author} {\bibfnamefont {H.}~\bibnamefont
  {Kim}}, \bibinfo {author} {\bibfnamefont {K.}~\bibnamefont {Watanabe}},
  \bibinfo {author} {\bibfnamefont {T.}~\bibnamefont {Taniguchi}}, \bibinfo
  {author} {\bibfnamefont {J.}~\bibnamefont {Alicea}}, \emph {et~al.},\
  }\bibfield  {title} {\bibinfo {title} {Ascendance of superconductivity in
  magic-angle graphene multilayers},\ }\href@noop {} {\bibfield  {journal}
  {\bibinfo  {journal} {arXiv preprint arXiv:2112.09270}\ } (\bibinfo {year}
  {2021})}\BibitemShut {NoStop}%
\bibitem [{\citenamefont {Park}\ \emph {et~al.}(2021)\citenamefont {Park},
  \citenamefont {Cao}, \citenamefont {Watanabe}, \citenamefont {Taniguchi},\
  and\ \citenamefont {Jarillo-Herrero}}]{park2021tunable}%
  \BibitemOpen
  \bibfield  {author} {\bibinfo {author} {\bibfnamefont {J.~M.}\ \bibnamefont
  {Park}}, \bibinfo {author} {\bibfnamefont {Y.}~\bibnamefont {Cao}}, \bibinfo
  {author} {\bibfnamefont {K.}~\bibnamefont {Watanabe}}, \bibinfo {author}
  {\bibfnamefont {T.}~\bibnamefont {Taniguchi}},\ and\ \bibinfo {author}
  {\bibfnamefont {P.}~\bibnamefont {Jarillo-Herrero}},\ }\bibfield  {title}
  {\bibinfo {title} {Tunable strongly coupled superconductivity in magic-angle
  twisted trilayer graphene},\ }\href@noop {} {\bibfield  {journal} {\bibinfo
  {journal} {Nature}\ }\textbf {\bibinfo {volume} {590}},\ \bibinfo {pages}
  {249} (\bibinfo {year} {2021})}\BibitemShut {NoStop}%
\bibitem [{\citenamefont {Cao}\ \emph {et~al.}(2021{\natexlab{b}})\citenamefont
  {Cao}, \citenamefont {Park}, \citenamefont {Watanabe}, \citenamefont
  {Taniguchi},\ and\ \citenamefont {Jarillo-Herrero}}]{cao2021pauli}%
  \BibitemOpen
  \bibfield  {author} {\bibinfo {author} {\bibfnamefont {Y.}~\bibnamefont
  {Cao}}, \bibinfo {author} {\bibfnamefont {J.~M.}\ \bibnamefont {Park}},
  \bibinfo {author} {\bibfnamefont {K.}~\bibnamefont {Watanabe}}, \bibinfo
  {author} {\bibfnamefont {T.}~\bibnamefont {Taniguchi}},\ and\ \bibinfo
  {author} {\bibfnamefont {P.}~\bibnamefont {Jarillo-Herrero}},\ }\bibfield
  {title} {\bibinfo {title} {Pauli-limit violation and re-entrant
  superconductivity in moir{\'e} graphene},\ }\href@noop {} {\bibfield
  {journal} {\bibinfo  {journal} {Nature}\ }\textbf {\bibinfo {volume} {595}},\
  \bibinfo {pages} {526} (\bibinfo {year} {2021}{\natexlab{b}})}\BibitemShut
  {NoStop}%
\bibitem [{\citenamefont {Kim}\ \emph {et~al.}(2022)\citenamefont {Kim},
  \citenamefont {Choi}, \citenamefont {Lewandowski}, \citenamefont {Thomson},
  \citenamefont {Zhang}, \citenamefont {Polski}, \citenamefont {Watanabe},
  \citenamefont {Taniguchi}, \citenamefont {Alicea},\ and\ \citenamefont
  {Nadj-Perge}}]{kim2022evidence}%
  \BibitemOpen
  \bibfield  {author} {\bibinfo {author} {\bibfnamefont {H.}~\bibnamefont
  {Kim}}, \bibinfo {author} {\bibfnamefont {Y.}~\bibnamefont {Choi}}, \bibinfo
  {author} {\bibfnamefont {C.}~\bibnamefont {Lewandowski}}, \bibinfo {author}
  {\bibfnamefont {A.}~\bibnamefont {Thomson}}, \bibinfo {author} {\bibfnamefont
  {Y.}~\bibnamefont {Zhang}}, \bibinfo {author} {\bibfnamefont
  {R.}~\bibnamefont {Polski}}, \bibinfo {author} {\bibfnamefont
  {K.}~\bibnamefont {Watanabe}}, \bibinfo {author} {\bibfnamefont
  {T.}~\bibnamefont {Taniguchi}}, \bibinfo {author} {\bibfnamefont
  {J.}~\bibnamefont {Alicea}},\ and\ \bibinfo {author} {\bibfnamefont
  {S.}~\bibnamefont {Nadj-Perge}},\ }\bibfield  {title} {\bibinfo {title}
  {Evidence for unconventional superconductivity in twisted trilayer
  graphene},\ }\href@noop {} {\bibfield  {journal} {\bibinfo  {journal}
  {Nature}\ }\textbf {\bibinfo {volume} {606}},\ \bibinfo {pages} {494}
  (\bibinfo {year} {2022})}\BibitemShut {NoStop}%
\bibitem [{\citenamefont {Liu}\ \emph {et~al.}(2022)\citenamefont {Liu},
  \citenamefont {Zhang}, \citenamefont {Watanabe}, \citenamefont {Taniguchi},\
  and\ \citenamefont {Li}}]{liu2022isospin}%
  \BibitemOpen
  \bibfield  {author} {\bibinfo {author} {\bibfnamefont {X.}~\bibnamefont
  {Liu}}, \bibinfo {author} {\bibfnamefont {N.~J.}\ \bibnamefont {Zhang}},
  \bibinfo {author} {\bibfnamefont {K.}~\bibnamefont {Watanabe}}, \bibinfo
  {author} {\bibfnamefont {T.}~\bibnamefont {Taniguchi}},\ and\ \bibinfo
  {author} {\bibfnamefont {J.}~\bibnamefont {Li}},\ }\bibfield  {title}
  {\bibinfo {title} {Isospin order in superconducting magic-angle twisted
  trilayer graphene},\ }\href@noop {} {\bibfield  {journal} {\bibinfo
  {journal} {Nature Physics}\ }\textbf {\bibinfo {volume} {18}},\ \bibinfo
  {pages} {522} (\bibinfo {year} {2022})}\BibitemShut {NoStop}%
\bibitem [{\citenamefont {Kerelsky}\ \emph {et~al.}(2019)\citenamefont
  {Kerelsky}, \citenamefont {McGilly}, \citenamefont {Kennes}, \citenamefont
  {Xian}, \citenamefont {Yankowitz}, \citenamefont {Chen}, \citenamefont
  {Watanabe}, \citenamefont {Taniguchi}, \citenamefont {Hone}, \citenamefont
  {Dean} \emph {et~al.}}]{kerelsky2019maximized}%
  \BibitemOpen
  \bibfield  {author} {\bibinfo {author} {\bibfnamefont {A.}~\bibnamefont
  {Kerelsky}}, \bibinfo {author} {\bibfnamefont {L.~J.}\ \bibnamefont
  {McGilly}}, \bibinfo {author} {\bibfnamefont {D.~M.}\ \bibnamefont {Kennes}},
  \bibinfo {author} {\bibfnamefont {L.}~\bibnamefont {Xian}}, \bibinfo {author}
  {\bibfnamefont {M.}~\bibnamefont {Yankowitz}}, \bibinfo {author}
  {\bibfnamefont {S.}~\bibnamefont {Chen}}, \bibinfo {author} {\bibfnamefont
  {K.}~\bibnamefont {Watanabe}}, \bibinfo {author} {\bibfnamefont
  {T.}~\bibnamefont {Taniguchi}}, \bibinfo {author} {\bibfnamefont
  {J.}~\bibnamefont {Hone}}, \bibinfo {author} {\bibfnamefont {C.}~\bibnamefont
  {Dean}}, \emph {et~al.},\ }\bibfield  {title} {\bibinfo {title} {Maximized
  electron interactions at the magic angle in twisted bilayer graphene},\
  }\href@noop {} {\bibfield  {journal} {\bibinfo  {journal} {Nature}\ }\textbf
  {\bibinfo {volume} {572}},\ \bibinfo {pages} {95} (\bibinfo {year}
  {2019})}\BibitemShut {NoStop}%
\bibitem [{\citenamefont {Balents}\ \emph {et~al.}(2020)\citenamefont
  {Balents}, \citenamefont {Dean}, \citenamefont {Efetov},\ and\ \citenamefont
  {Young}}]{Balents2020}%
  \BibitemOpen
  \bibfield  {author} {\bibinfo {author} {\bibfnamefont {L.}~\bibnamefont
  {Balents}}, \bibinfo {author} {\bibfnamefont {C.~R.}\ \bibnamefont {Dean}},
  \bibinfo {author} {\bibfnamefont {D.~K.}\ \bibnamefont {Efetov}},\ and\
  \bibinfo {author} {\bibfnamefont {A.~F.}\ \bibnamefont {Young}},\ }\bibfield
  {title} {\bibinfo {title} {Superconductivity and strong correlations in
  moir{\'e} flat bands},\ }\href {https://doi.org/10.1038/s41567-020-0906-9}
  {\bibfield  {journal} {\bibinfo  {journal} {Nature Physics}\ }\textbf
  {\bibinfo {volume} {16}},\ \bibinfo {pages} {725} (\bibinfo {year}
  {2020})}\BibitemShut {NoStop}%
\bibitem [{\citenamefont {Jung}\ and\ \citenamefont
  {MacDonald}(2014)}]{jung2014accurate}%
  \BibitemOpen
  \bibfield  {author} {\bibinfo {author} {\bibfnamefont {J.}~\bibnamefont
  {Jung}}\ and\ \bibinfo {author} {\bibfnamefont {A.~H.}\ \bibnamefont
  {MacDonald}},\ }\bibfield  {title} {\bibinfo {title} {Accurate tight-binding
  models for the $\pi$ bands of bilayer graphene},\ }\href@noop {} {\bibfield
  {journal} {\bibinfo  {journal} {Physical Review B}\ }\textbf {\bibinfo
  {volume} {89}},\ \bibinfo {pages} {035405} (\bibinfo {year}
  {2014})}\BibitemShut {NoStop}%
\bibitem [{\citenamefont {Rozhkov}\ \emph {et~al.}(2016)\citenamefont
  {Rozhkov}, \citenamefont {Sboychakov}, \citenamefont {Rakhmanov},\ and\
  \citenamefont {Nori}}]{rozhkov2016electronic}%
  \BibitemOpen
  \bibfield  {author} {\bibinfo {author} {\bibfnamefont {A.~V.}\ \bibnamefont
  {Rozhkov}}, \bibinfo {author} {\bibfnamefont {A.}~\bibnamefont {Sboychakov}},
  \bibinfo {author} {\bibfnamefont {A.}~\bibnamefont {Rakhmanov}},\ and\
  \bibinfo {author} {\bibfnamefont {F.}~\bibnamefont {Nori}},\ }\bibfield
  {title} {\bibinfo {title} {Electronic properties of graphene-based bilayer
  systems},\ }\href@noop {} {\bibfield  {journal} {\bibinfo  {journal} {Physics
  Reports}\ }\textbf {\bibinfo {volume} {648}},\ \bibinfo {pages} {1} (\bibinfo
  {year} {2016})}\BibitemShut {NoStop}%
\bibitem [{\citenamefont {Gmitra}\ \emph {et~al.}(2009)\citenamefont {Gmitra},
  \citenamefont {Konschuh}, \citenamefont {Ertler}, \citenamefont
  {Ambrosch-Draxl},\ and\ \citenamefont {Fabian}}]{gmitra2009band}%
  \BibitemOpen
  \bibfield  {author} {\bibinfo {author} {\bibfnamefont {M.}~\bibnamefont
  {Gmitra}}, \bibinfo {author} {\bibfnamefont {S.}~\bibnamefont {Konschuh}},
  \bibinfo {author} {\bibfnamefont {C.}~\bibnamefont {Ertler}}, \bibinfo
  {author} {\bibfnamefont {C.}~\bibnamefont {Ambrosch-Draxl}},\ and\ \bibinfo
  {author} {\bibfnamefont {J.}~\bibnamefont {Fabian}},\ }\bibfield  {title}
  {\bibinfo {title} {Band-structure topologies of graphene: Spin-orbit coupling
  effects from first principles},\ }\href@noop {} {\bibfield  {journal}
  {\bibinfo  {journal} {Physical Review B}\ }\textbf {\bibinfo {volume} {80}},\
  \bibinfo {pages} {235431} (\bibinfo {year} {2009})}\BibitemShut {NoStop}%
\bibitem [{\citenamefont {Gmitra}\ and\ \citenamefont
  {Fabian}(2015)}]{gmitra2015graphene}%
  \BibitemOpen
  \bibfield  {author} {\bibinfo {author} {\bibfnamefont {M.}~\bibnamefont
  {Gmitra}}\ and\ \bibinfo {author} {\bibfnamefont {J.}~\bibnamefont
  {Fabian}},\ }\bibfield  {title} {\bibinfo {title} {Graphene on
  transition-metal dichalcogenides: A platform for proximity spin-orbit physics
  and optospintronics},\ }\href@noop {} {\bibfield  {journal} {\bibinfo
  {journal} {Physical Review B}\ }\textbf {\bibinfo {volume} {92}},\ \bibinfo
  {pages} {155403} (\bibinfo {year} {2015})}\BibitemShut {NoStop}%
\bibitem [{\citenamefont {Zollner}\ \emph {et~al.}(2016)\citenamefont
  {Zollner}, \citenamefont {Gmitra}, \citenamefont {Frank},\ and\ \citenamefont
  {Fabian}}]{zollner2016theory}%
  \BibitemOpen
  \bibfield  {author} {\bibinfo {author} {\bibfnamefont {K.}~\bibnamefont
  {Zollner}}, \bibinfo {author} {\bibfnamefont {M.}~\bibnamefont {Gmitra}},
  \bibinfo {author} {\bibfnamefont {T.}~\bibnamefont {Frank}},\ and\ \bibinfo
  {author} {\bibfnamefont {J.}~\bibnamefont {Fabian}},\ }\bibfield  {title}
  {\bibinfo {title} {Theory of proximity-induced exchange coupling in graphene
  on hbn/(co, ni)},\ }\href@noop {} {\bibfield  {journal} {\bibinfo  {journal}
  {Physical Review B}\ }\textbf {\bibinfo {volume} {94}},\ \bibinfo {pages}
  {155441} (\bibinfo {year} {2016})}\BibitemShut {NoStop}%
\bibitem [{\citenamefont {Sierra}\ \emph {et~al.}(2021)\citenamefont {Sierra},
  \citenamefont {Fabian}, \citenamefont {Kawakami}, \citenamefont {Roche},\
  and\ \citenamefont {Valenzuela}}]{sierra2021van}%
  \BibitemOpen
  \bibfield  {author} {\bibinfo {author} {\bibfnamefont {J.~F.}\ \bibnamefont
  {Sierra}}, \bibinfo {author} {\bibfnamefont {J.}~\bibnamefont {Fabian}},
  \bibinfo {author} {\bibfnamefont {R.~K.}\ \bibnamefont {Kawakami}}, \bibinfo
  {author} {\bibfnamefont {S.}~\bibnamefont {Roche}},\ and\ \bibinfo {author}
  {\bibfnamefont {S.~O.}\ \bibnamefont {Valenzuela}},\ }\bibfield  {title}
  {\bibinfo {title} {Van der waals heterostructures for spintronics and
  opto-spintronics},\ }\href@noop {} {\bibfield  {journal} {\bibinfo  {journal}
  {Nature Nanotechnology}\ }\textbf {\bibinfo {volume} {16}},\ \bibinfo {pages}
  {856} (\bibinfo {year} {2021})}\BibitemShut {NoStop}%
\bibitem [{\citenamefont {Metzner}\ \emph {et~al.}(2012)\citenamefont
  {Metzner}, \citenamefont {Salmhofer}, \citenamefont {Honerkamp},
  \citenamefont {Meden},\ and\ \citenamefont
  {Sch\"onhammer}}]{metzner2012functional}%
  \BibitemOpen
  \bibfield  {author} {\bibinfo {author} {\bibfnamefont {W.}~\bibnamefont
  {Metzner}}, \bibinfo {author} {\bibfnamefont {M.}~\bibnamefont {Salmhofer}},
  \bibinfo {author} {\bibfnamefont {C.}~\bibnamefont {Honerkamp}}, \bibinfo
  {author} {\bibfnamefont {V.}~\bibnamefont {Meden}},\ and\ \bibinfo {author}
  {\bibfnamefont {K.}~\bibnamefont {Sch\"onhammer}},\ }\bibfield  {title}
  {\bibinfo {title} {Functional renormalization group approach to correlated
  fermion systems},\ }\href {https://doi.org/10.1103/RevModPhys.84.299}
  {\bibfield  {journal} {\bibinfo  {journal} {Rev. Mod. Phys.}\ }\textbf
  {\bibinfo {volume} {84}},\ \bibinfo {pages} {299} (\bibinfo {year}
  {2012})}\BibitemShut {NoStop}%
\bibitem [{\citenamefont {Platt}\ \emph {et~al.}(2013)\citenamefont {Platt},
  \citenamefont {Hanke},\ and\ \citenamefont {Thomale}}]{Platt2013}%
  \BibitemOpen
  \bibfield  {author} {\bibinfo {author} {\bibfnamefont {C.}~\bibnamefont
  {Platt}}, \bibinfo {author} {\bibfnamefont {W.}~\bibnamefont {Hanke}},\ and\
  \bibinfo {author} {\bibfnamefont {R.}~\bibnamefont {Thomale}},\ }\bibfield
  {title} {\bibinfo {title} {Functional renormalization group for multi-orbital
  fermi surface instabilities},\ }\href
  {https://doi.org/10.1080/00018732.2013.862020} {\bibfield  {journal}
  {\bibinfo  {journal} {Advances in Physics}\ }\textbf {\bibinfo {volume}
  {62}},\ \bibinfo {pages} {453} (\bibinfo {year} {2013})}\BibitemShut
  {NoStop}%
\bibitem [{\citenamefont {Honerkamp}(2008)}]{Honerkamp2008}%
  \BibitemOpen
  \bibfield  {author} {\bibinfo {author} {\bibfnamefont {C.}~\bibnamefont
  {Honerkamp}},\ }\bibfield  {title} {\bibinfo {title} {Density waves and
  cooper pairing on the honeycomb lattice},\ }\href
  {https://doi.org/10.1103/PhysRevLett.100.146404} {\bibfield  {journal}
  {\bibinfo  {journal} {Phys. Rev. Lett.}\ }\textbf {\bibinfo {volume} {100}},\
  \bibinfo {pages} {146404} (\bibinfo {year} {2008})}\BibitemShut {NoStop}%
\bibitem [{\citenamefont {Georges}\ \emph {et~al.}(1996)\citenamefont
  {Georges}, \citenamefont {Kotliar}, \citenamefont {Krauth},\ and\
  \citenamefont {Rozenberg}}]{georges1996dynamical}%
  \BibitemOpen
  \bibfield  {author} {\bibinfo {author} {\bibfnamefont {A.}~\bibnamefont
  {Georges}}, \bibinfo {author} {\bibfnamefont {G.}~\bibnamefont {Kotliar}},
  \bibinfo {author} {\bibfnamefont {W.}~\bibnamefont {Krauth}},\ and\ \bibinfo
  {author} {\bibfnamefont {M.~J.}\ \bibnamefont {Rozenberg}},\ }\bibfield
  {title} {\bibinfo {title} {Dynamical mean-field theory of strongly correlated
  fermion systems and the limit of infinite dimensions},\ }\href@noop {}
  {\bibfield  {journal} {\bibinfo  {journal} {Reviews of Modern Physics}\
  }\textbf {\bibinfo {volume} {68}},\ \bibinfo {pages} {13} (\bibinfo {year}
  {1996})}\BibitemShut {NoStop}%
\bibitem [{\citenamefont {Schollw{\"o}ck}(2011)}]{schollwock2011density}%
  \BibitemOpen
  \bibfield  {author} {\bibinfo {author} {\bibfnamefont {U.}~\bibnamefont
  {Schollw{\"o}ck}},\ }\bibfield  {title} {\bibinfo {title} {The density-matrix
  renormalization group in the age of matrix product states},\ }\href@noop {}
  {\bibfield  {journal} {\bibinfo  {journal} {Annals of physics}\ }\textbf
  {\bibinfo {volume} {326}},\ \bibinfo {pages} {96} (\bibinfo {year}
  {2011})}\BibitemShut {NoStop}%
\bibitem [{\citenamefont {Or{\'u}s}(2014)}]{orus2014practical}%
  \BibitemOpen
  \bibfield  {author} {\bibinfo {author} {\bibfnamefont {R.}~\bibnamefont
  {Or{\'u}s}},\ }\bibfield  {title} {\bibinfo {title} {A practical introduction
  to tensor networks: Matrix product states and projected entangled pair
  states},\ }\href@noop {} {\bibfield  {journal} {\bibinfo  {journal} {Annals
  of physics}\ }\textbf {\bibinfo {volume} {349}},\ \bibinfo {pages} {117}
  (\bibinfo {year} {2014})}\BibitemShut {NoStop}%
\bibitem [{\citenamefont {Banszerus}\ \emph {et~al.}(2023)\citenamefont
  {Banszerus}, \citenamefont {M{\"o}ller}, \citenamefont {Hecker},
  \citenamefont {Icking}, \citenamefont {Watanabe}, \citenamefont {Taniguchi},
  \citenamefont {Hassler}, \citenamefont {Volk},\ and\ \citenamefont
  {Stampfer}}]{banszerus2023particle}%
  \BibitemOpen
  \bibfield  {author} {\bibinfo {author} {\bibfnamefont {L.}~\bibnamefont
  {Banszerus}}, \bibinfo {author} {\bibfnamefont {S.}~\bibnamefont
  {M{\"o}ller}}, \bibinfo {author} {\bibfnamefont {K.}~\bibnamefont {Hecker}},
  \bibinfo {author} {\bibfnamefont {E.}~\bibnamefont {Icking}}, \bibinfo
  {author} {\bibfnamefont {K.}~\bibnamefont {Watanabe}}, \bibinfo {author}
  {\bibfnamefont {T.}~\bibnamefont {Taniguchi}}, \bibinfo {author}
  {\bibfnamefont {F.}~\bibnamefont {Hassler}}, \bibinfo {author} {\bibfnamefont
  {C.}~\bibnamefont {Volk}},\ and\ \bibinfo {author} {\bibfnamefont
  {C.}~\bibnamefont {Stampfer}},\ }\bibfield  {title} {\bibinfo {title}
  {Particle--hole symmetry protects spin-valley blockade in graphene quantum
  dots},\ }\href@noop {} {\bibfield  {journal} {\bibinfo  {journal} {Nature}\
  ,\ \bibinfo {pages} {1}} (\bibinfo {year} {2023})}\BibitemShut {NoStop}%
\bibitem [{\citenamefont {Wirth}\ \emph {et~al.}(2022)\citenamefont {Wirth},
  \citenamefont {Profe}, \citenamefont {Rothstein}, \citenamefont {Kyoseva},
  \citenamefont {Siebenkotten}, \citenamefont {Conrads}, \citenamefont {Klebl},
  \citenamefont {Fischer}, \citenamefont {Beschoten}, \citenamefont {Stampfer},
  \citenamefont {Kennes}, \citenamefont {Waldecker},\ and\ \citenamefont
  {Taubner}}]{Wirth2022Oct}%
  \BibitemOpen
  \bibfield  {author} {\bibinfo {author} {\bibfnamefont {K.~G.}\ \bibnamefont
  {Wirth}}, \bibinfo {author} {\bibfnamefont {J.~B.}\ \bibnamefont {Profe}},
  \bibinfo {author} {\bibfnamefont {A.}~\bibnamefont {Rothstein}}, \bibinfo
  {author} {\bibfnamefont {H.}~\bibnamefont {Kyoseva}}, \bibinfo {author}
  {\bibfnamefont {D.}~\bibnamefont {Siebenkotten}}, \bibinfo {author}
  {\bibfnamefont {L.}~\bibnamefont {Conrads}}, \bibinfo {author} {\bibfnamefont
  {L.}~\bibnamefont {Klebl}}, \bibinfo {author} {\bibfnamefont
  {A.}~\bibnamefont {Fischer}}, \bibinfo {author} {\bibfnamefont
  {B.}~\bibnamefont {Beschoten}}, \bibinfo {author} {\bibfnamefont
  {C.}~\bibnamefont {Stampfer}}, \bibinfo {author} {\bibfnamefont {D.~M.}\
  \bibnamefont {Kennes}}, \bibinfo {author} {\bibfnamefont {L.}~\bibnamefont
  {Waldecker}},\ and\ \bibinfo {author} {\bibfnamefont {T.}~\bibnamefont
  {Taubner}},\ }\bibfield  {title} {\bibinfo {title} {{Experimental Observation
  of ABCB Stacked Tetralayer Graphene}},\ }\href
  {https://doi.org/10.1021/acsnano.2c06053} {\bibfield  {journal} {\bibinfo
  {journal} {ACS Nano}\ }\textbf {\bibinfo {volume} {16}},\ \bibinfo {pages}
  {16617} (\bibinfo {year} {2022})}\BibitemShut {NoStop}%
\bibitem [{\citenamefont {Atri}\ \emph {et~al.}(2023)\citenamefont {Atri},
  \citenamefont {Cao}, \citenamefont {Alon}, \citenamefont {Roy}, \citenamefont
  {Stern}, \citenamefont {Falko}, \citenamefont {Goldstein}, \citenamefont
  {Kronik}, \citenamefont {Urbakh}, \citenamefont {Hod},\ and\ \citenamefont
  {Shalom}}]{atri2023spontaneous}%
  \BibitemOpen
  \bibfield  {author} {\bibinfo {author} {\bibfnamefont {S.~S.}\ \bibnamefont
  {Atri}}, \bibinfo {author} {\bibfnamefont {W.}~\bibnamefont {Cao}}, \bibinfo
  {author} {\bibfnamefont {B.}~\bibnamefont {Alon}}, \bibinfo {author}
  {\bibfnamefont {N.}~\bibnamefont {Roy}}, \bibinfo {author} {\bibfnamefont
  {M.~V.}\ \bibnamefont {Stern}}, \bibinfo {author} {\bibfnamefont
  {V.}~\bibnamefont {Falko}}, \bibinfo {author} {\bibfnamefont
  {M.}~\bibnamefont {Goldstein}}, \bibinfo {author} {\bibfnamefont
  {L.}~\bibnamefont {Kronik}}, \bibinfo {author} {\bibfnamefont
  {M.}~\bibnamefont {Urbakh}}, \bibinfo {author} {\bibfnamefont
  {O.}~\bibnamefont {Hod}},\ and\ \bibinfo {author} {\bibfnamefont {M.~B.}\
  \bibnamefont {Shalom}},\ }\href@noop {} {\bibinfo {title} {Spontaneous
  electric polarization in graphene polytypes}} (\bibinfo {year} {2023}),\
  \Eprint {https://arxiv.org/abs/2305.10890} {arXiv:2305.10890
  [cond-mat.mtrl-sci]} \BibitemShut {NoStop}%
\bibitem [{\citenamefont {Hossain}\ \emph {et~al.}(2021)\citenamefont
  {Hossain}, \citenamefont {Ma}, \citenamefont {Villegas-Rosales},
  \citenamefont {Chung}, \citenamefont {Pfeiffer}, \citenamefont {West},
  \citenamefont {Baldwin},\ and\ \citenamefont {Shayegan}}]{hossain2021}%
  \BibitemOpen
  \bibfield  {author} {\bibinfo {author} {\bibfnamefont {M.~S.}\ \bibnamefont
  {Hossain}}, \bibinfo {author} {\bibfnamefont {M.~K.}\ \bibnamefont {Ma}},
  \bibinfo {author} {\bibfnamefont {K.~A.}\ \bibnamefont {Villegas-Rosales}},
  \bibinfo {author} {\bibfnamefont {Y.~J.}\ \bibnamefont {Chung}}, \bibinfo
  {author} {\bibfnamefont {L.~N.}\ \bibnamefont {Pfeiffer}}, \bibinfo {author}
  {\bibfnamefont {K.~W.}\ \bibnamefont {West}}, \bibinfo {author}
  {\bibfnamefont {K.~W.}\ \bibnamefont {Baldwin}},\ and\ \bibinfo {author}
  {\bibfnamefont {M.}~\bibnamefont {Shayegan}},\ }\bibfield  {title} {\bibinfo
  {title} {Spontaneous valley polarization of itinerant electrons},\ }\href
  {https://doi.org/10.1103/PhysRevLett.127.116601} {\bibfield  {journal}
  {\bibinfo  {journal} {Phys. Rev. Lett.}\ }\textbf {\bibinfo {volume} {127}},\
  \bibinfo {pages} {116601} (\bibinfo {year} {2021})}\BibitemShut {NoStop}%
\bibitem [{\citenamefont {Shkolnikov}\ \emph {et~al.}(2002)\citenamefont
  {Shkolnikov}, \citenamefont {De~Poortere}, \citenamefont {Tutuc},\ and\
  \citenamefont {Shayegan}}]{shkolnikov2002}%
  \BibitemOpen
  \bibfield  {author} {\bibinfo {author} {\bibfnamefont {Y.~P.}\ \bibnamefont
  {Shkolnikov}}, \bibinfo {author} {\bibfnamefont {E.~P.}\ \bibnamefont
  {De~Poortere}}, \bibinfo {author} {\bibfnamefont {E.}~\bibnamefont {Tutuc}},\
  and\ \bibinfo {author} {\bibfnamefont {M.}~\bibnamefont {Shayegan}},\
  }\bibfield  {title} {\bibinfo {title} {Valley splitting of alas
  two-dimensional electrons in a perpendicular magnetic field},\ }\href
  {https://doi.org/10.1103/PhysRevLett.89.226805} {\bibfield  {journal}
  {\bibinfo  {journal} {Phys. Rev. Lett.}\ }\textbf {\bibinfo {volume} {89}},\
  \bibinfo {pages} {226805} (\bibinfo {year} {2002})}\BibitemShut {NoStop}%
\bibitem [{\citenamefont {Wehling}\ \emph {et~al.}(2011)\citenamefont
  {Wehling}, \citenamefont {\ifmmode \mbox{\c{S}}\else \c{S}\fi{}a\ifmmode
  \mbox{\c{s}}\else \c{s}\fi{}\ifmmode \imath \else \i
  \fi{}o\ifmmode~\breve{g}\else \u{g}\fi{}lu}, \citenamefont {Friedrich},
  \citenamefont {Lichtenstein}, \citenamefont {Katsnelson},\ and\ \citenamefont
  {Bl\"ugel}}]{Wehling2011}%
  \BibitemOpen
  \bibfield  {author} {\bibinfo {author} {\bibfnamefont {T.~O.}\ \bibnamefont
  {Wehling}}, \bibinfo {author} {\bibfnamefont {E.}~\bibnamefont {\ifmmode
  \mbox{\c{S}}\else \c{S}\fi{}a\ifmmode \mbox{\c{s}}\else \c{s}\fi{}\ifmmode
  \imath \else \i \fi{}o\ifmmode~\breve{g}\else \u{g}\fi{}lu}}, \bibinfo
  {author} {\bibfnamefont {C.}~\bibnamefont {Friedrich}}, \bibinfo {author}
  {\bibfnamefont {A.~I.}\ \bibnamefont {Lichtenstein}}, \bibinfo {author}
  {\bibfnamefont {M.~I.}\ \bibnamefont {Katsnelson}},\ and\ \bibinfo {author}
  {\bibfnamefont {S.}~\bibnamefont {Bl\"ugel}},\ }\bibfield  {title} {\bibinfo
  {title} {Strength of effective coulomb interactions in graphene and
  graphite},\ }\href {https://doi.org/10.1103/PhysRevLett.106.236805}
  {\bibfield  {journal} {\bibinfo  {journal} {Phys. Rev. Lett.}\ }\textbf
  {\bibinfo {volume} {106}},\ \bibinfo {pages} {236805} (\bibinfo {year}
  {2011})}\BibitemShut {NoStop}%
\bibitem [{\citenamefont {R{\"o}sner}\ \emph {et~al.}(2015)\citenamefont
  {R{\"o}sner}, \citenamefont {{\c{S}}a{\c{s}}{\i}o{\u{g}}lu}, \citenamefont
  {Friedrich}, \citenamefont {Bl{\"u}gel},\ and\ \citenamefont
  {Wehling}}]{rosner2015wannier}%
  \BibitemOpen
  \bibfield  {author} {\bibinfo {author} {\bibfnamefont {M.}~\bibnamefont
  {R{\"o}sner}}, \bibinfo {author} {\bibfnamefont {E.}~\bibnamefont
  {{\c{S}}a{\c{s}}{\i}o{\u{g}}lu}}, \bibinfo {author} {\bibfnamefont
  {C.}~\bibnamefont {Friedrich}}, \bibinfo {author} {\bibfnamefont
  {S.}~\bibnamefont {Bl{\"u}gel}},\ and\ \bibinfo {author} {\bibfnamefont
  {T.}~\bibnamefont {Wehling}},\ }\bibfield  {title} {\bibinfo {title} {Wannier
  function approach to realistic coulomb interactions in layered materials and
  heterostructures},\ }\href@noop {} {\bibfield  {journal} {\bibinfo  {journal}
  {Physical Review B}\ }\textbf {\bibinfo {volume} {92}},\ \bibinfo {pages}
  {085102} (\bibinfo {year} {2015})}\BibitemShut {NoStop}%
\bibitem [{\citenamefont {Bistritzer}\ and\ \citenamefont
  {MacDonald}(2011)}]{bistritzer2011moire}%
  \BibitemOpen
  \bibfield  {author} {\bibinfo {author} {\bibfnamefont {R.}~\bibnamefont
  {Bistritzer}}\ and\ \bibinfo {author} {\bibfnamefont {A.~H.}\ \bibnamefont
  {MacDonald}},\ }\bibfield  {title} {\bibinfo {title} {Moir{\'e} bands in
  twisted double-layer graphene},\ }\href@noop {} {\bibfield  {journal}
  {\bibinfo  {journal} {Proceedings of the National Academy of Sciences}\
  }\textbf {\bibinfo {volume} {108}},\ \bibinfo {pages} {12233} (\bibinfo
  {year} {2011})}\BibitemShut {NoStop}%
\bibitem [{\citenamefont {Dos~Santos}\ \emph {et~al.}(2007)\citenamefont
  {Dos~Santos}, \citenamefont {Peres},\ and\ \citenamefont
  {Neto}}]{dos2007graphene}%
  \BibitemOpen
  \bibfield  {author} {\bibinfo {author} {\bibfnamefont {J.~L.}\ \bibnamefont
  {Dos~Santos}}, \bibinfo {author} {\bibfnamefont {N.}~\bibnamefont {Peres}},\
  and\ \bibinfo {author} {\bibfnamefont {A.~C.}\ \bibnamefont {Neto}},\
  }\bibfield  {title} {\bibinfo {title} {Graphene bilayer with a twist:
  Electronic structure},\ }\href@noop {} {\bibfield  {journal} {\bibinfo
  {journal} {Physical review letters}\ }\textbf {\bibinfo {volume} {99}},\
  \bibinfo {pages} {256802} (\bibinfo {year} {2007})}\BibitemShut {NoStop}%
\bibitem [{\citenamefont {Shallcross}\ \emph {et~al.}(2010)\citenamefont
  {Shallcross}, \citenamefont {Sharma}, \citenamefont {Kandelaki},\ and\
  \citenamefont {Pankratov}}]{shallcross2010electronic}%
  \BibitemOpen
  \bibfield  {author} {\bibinfo {author} {\bibfnamefont {S.}~\bibnamefont
  {Shallcross}}, \bibinfo {author} {\bibfnamefont {S.}~\bibnamefont {Sharma}},
  \bibinfo {author} {\bibfnamefont {E.}~\bibnamefont {Kandelaki}},\ and\
  \bibinfo {author} {\bibfnamefont {O.}~\bibnamefont {Pankratov}},\ }\bibfield
  {title} {\bibinfo {title} {Electronic structure of turbostratic graphene},\
  }\href@noop {} {\bibfield  {journal} {\bibinfo  {journal} {Physical Review
  B}\ }\textbf {\bibinfo {volume} {81}},\ \bibinfo {pages} {165105} (\bibinfo
  {year} {2010})}\BibitemShut {NoStop}%
\bibitem [{\citenamefont {Morell}\ \emph {et~al.}(2010)\citenamefont {Morell},
  \citenamefont {Correa}, \citenamefont {Vargas}, \citenamefont {Pacheco},\
  and\ \citenamefont {Barticevic}}]{morell2010flat}%
  \BibitemOpen
  \bibfield  {author} {\bibinfo {author} {\bibfnamefont {E.~S.}\ \bibnamefont
  {Morell}}, \bibinfo {author} {\bibfnamefont {J.}~\bibnamefont {Correa}},
  \bibinfo {author} {\bibfnamefont {P.}~\bibnamefont {Vargas}}, \bibinfo
  {author} {\bibfnamefont {M.}~\bibnamefont {Pacheco}},\ and\ \bibinfo {author}
  {\bibfnamefont {Z.}~\bibnamefont {Barticevic}},\ }\bibfield  {title}
  {\bibinfo {title} {Flat bands in slightly twisted bilayer graphene:
  Tight-binding calculations},\ }\href@noop {} {\bibfield  {journal} {\bibinfo
  {journal} {Physical Review B}\ }\textbf {\bibinfo {volume} {82}},\ \bibinfo
  {pages} {121407} (\bibinfo {year} {2010})}\BibitemShut {NoStop}%
\bibitem [{\citenamefont {Carr}\ \emph
  {et~al.}(2019{\natexlab{a}})\citenamefont {Carr}, \citenamefont {Fang},
  \citenamefont {Zhu},\ and\ \citenamefont {Kaxiras}}]{carr2019relaxation}%
  \BibitemOpen
  \bibfield  {author} {\bibinfo {author} {\bibfnamefont {S.}~\bibnamefont
  {Carr}}, \bibinfo {author} {\bibfnamefont {S.}~\bibnamefont {Fang}}, \bibinfo
  {author} {\bibfnamefont {Z.}~\bibnamefont {Zhu}},\ and\ \bibinfo {author}
  {\bibfnamefont {E.}~\bibnamefont {Kaxiras}},\ }\bibfield  {title} {\bibinfo
  {title} {Exact continuum model for low-energy electronic states of twisted
  bilayer graphene},\ }\href {https://doi.org/10.1103/PhysRevResearch.1.013001}
  {\bibfield  {journal} {\bibinfo  {journal} {Phys. Rev. Res.}\ }\textbf
  {\bibinfo {volume} {1}},\ \bibinfo {pages} {013001} (\bibinfo {year}
  {2019}{\natexlab{a}})}\BibitemShut {NoStop}%
\bibitem [{\citenamefont {Carr}\ \emph
  {et~al.}(2019{\natexlab{b}})\citenamefont {Carr}, \citenamefont {Fang},
  \citenamefont {Po}, \citenamefont {Vishwanath},\ and\ \citenamefont
  {Kaxiras}}]{carr2019wannier}%
  \BibitemOpen
  \bibfield  {author} {\bibinfo {author} {\bibfnamefont {S.}~\bibnamefont
  {Carr}}, \bibinfo {author} {\bibfnamefont {S.}~\bibnamefont {Fang}}, \bibinfo
  {author} {\bibfnamefont {H.~C.}\ \bibnamefont {Po}}, \bibinfo {author}
  {\bibfnamefont {A.}~\bibnamefont {Vishwanath}},\ and\ \bibinfo {author}
  {\bibfnamefont {E.}~\bibnamefont {Kaxiras}},\ }\bibfield  {title} {\bibinfo
  {title} {Derivation of wannier orbitals and minimal-basis tight-binding
  hamiltonians for twisted bilayer graphene: First-principles approach},\
  }\href {https://doi.org/10.1103/PhysRevResearch.1.033072} {\bibfield
  {journal} {\bibinfo  {journal} {Phys. Rev. Res.}\ }\textbf {\bibinfo {volume}
  {1}},\ \bibinfo {pages} {033072} (\bibinfo {year}
  {2019}{\natexlab{b}})}\BibitemShut {NoStop}%
\bibitem [{\citenamefont {Song}\ and\ \citenamefont
  {Bernevig}(2022)}]{song2022magic}%
  \BibitemOpen
  \bibfield  {author} {\bibinfo {author} {\bibfnamefont {Z.-D.}\ \bibnamefont
  {Song}}\ and\ \bibinfo {author} {\bibfnamefont {B.~A.}\ \bibnamefont
  {Bernevig}},\ }\bibfield  {title} {\bibinfo {title} {Magic-angle twisted
  bilayer graphene as a topological heavy fermion problem},\ }\href@noop {}
  {\bibfield  {journal} {\bibinfo  {journal} {Physical review letters}\
  }\textbf {\bibinfo {volume} {129}},\ \bibinfo {pages} {047601} (\bibinfo
  {year} {2022})}\BibitemShut {NoStop}%
\bibitem [{\citenamefont {Hu}\ \emph {et~al.}(2023)\citenamefont {Hu},
  \citenamefont {Rai}, \citenamefont {Crippa}, \citenamefont
  {Herzog-Arbeitman}, \citenamefont {C\ifmmode \u{a}\else
  \u{a}\fi{}lug\ifmmode~\u{a}\else \u{a}\fi{}ru}, \citenamefont {Wehling},
  \citenamefont {Sangiovanni}, \citenamefont {Valent\'{\i}}, \citenamefont
  {Tsvelik},\ and\ \citenamefont {Bernevig}}]{hu2023symmetric}%
  \BibitemOpen
  \bibfield  {author} {\bibinfo {author} {\bibfnamefont {H.}~\bibnamefont
  {Hu}}, \bibinfo {author} {\bibfnamefont {G.}~\bibnamefont {Rai}}, \bibinfo
  {author} {\bibfnamefont {L.}~\bibnamefont {Crippa}}, \bibinfo {author}
  {\bibfnamefont {J.}~\bibnamefont {Herzog-Arbeitman}}, \bibinfo {author}
  {\bibfnamefont {D.}~\bibnamefont {C\ifmmode \u{a}\else
  \u{a}\fi{}lug\ifmmode~\u{a}\else \u{a}\fi{}ru}}, \bibinfo {author}
  {\bibfnamefont {T.}~\bibnamefont {Wehling}}, \bibinfo {author} {\bibfnamefont
  {G.}~\bibnamefont {Sangiovanni}}, \bibinfo {author} {\bibfnamefont
  {R.}~\bibnamefont {Valent\'{\i}}}, \bibinfo {author} {\bibfnamefont {A.~M.}\
  \bibnamefont {Tsvelik}},\ and\ \bibinfo {author} {\bibfnamefont {B.~A.}\
  \bibnamefont {Bernevig}},\ }\bibfield  {title} {\bibinfo {title} {Symmetric
  kondo lattice states in doped strained twisted bilayer graphene},\ }\href
  {https://doi.org/10.1103/PhysRevLett.131.166501} {\bibfield  {journal}
  {\bibinfo  {journal} {Phys. Rev. Lett.}\ }\textbf {\bibinfo {volume} {131}},\
  \bibinfo {pages} {166501} (\bibinfo {year} {2023})}\BibitemShut {NoStop}%
\bibitem [{\citenamefont {Yu}\ \emph {et~al.}(2023)\citenamefont {Yu},
  \citenamefont {Xie}, \citenamefont {Bernevig},\ and\ \citenamefont
  {Das~Sarma}}]{yu2023magic}%
  \BibitemOpen
  \bibfield  {author} {\bibinfo {author} {\bibfnamefont {J.}~\bibnamefont
  {Yu}}, \bibinfo {author} {\bibfnamefont {M.}~\bibnamefont {Xie}}, \bibinfo
  {author} {\bibfnamefont {B.~A.}\ \bibnamefont {Bernevig}},\ and\ \bibinfo
  {author} {\bibfnamefont {S.}~\bibnamefont {Das~Sarma}},\ }\bibfield  {title}
  {\bibinfo {title} {Magic-angle twisted symmetric trilayer graphene as a
  topological heavy-fermion problem},\ }\href
  {https://doi.org/10.1103/PhysRevB.108.035129} {\bibfield  {journal} {\bibinfo
   {journal} {Phys. Rev. B}\ }\textbf {\bibinfo {volume} {108}},\ \bibinfo
  {pages} {035129} (\bibinfo {year} {2023})}\BibitemShut {NoStop}%
\bibitem [{\citenamefont {Rai}\ \emph {et~al.}(2023)\citenamefont {Rai},
  \citenamefont {Crippa}, \citenamefont {Călugăru}, \citenamefont {Hu},
  \citenamefont {de' Medici}, \citenamefont {Georges}, \citenamefont
  {Bernevig}, \citenamefont {Valentí}, \citenamefont {Sangiovanni},\ and\
  \citenamefont {Wehling}}]{rai2023dynamicalcorrelationsordermagicangle}%
  \BibitemOpen
  \bibfield  {author} {\bibinfo {author} {\bibfnamefont {G.}~\bibnamefont
  {Rai}}, \bibinfo {author} {\bibfnamefont {L.}~\bibnamefont {Crippa}},
  \bibinfo {author} {\bibfnamefont {D.}~\bibnamefont {Călugăru}}, \bibinfo
  {author} {\bibfnamefont {H.}~\bibnamefont {Hu}}, \bibinfo {author}
  {\bibfnamefont {L.}~\bibnamefont {de' Medici}}, \bibinfo {author}
  {\bibfnamefont {A.}~\bibnamefont {Georges}}, \bibinfo {author} {\bibfnamefont
  {B.~A.}\ \bibnamefont {Bernevig}}, \bibinfo {author} {\bibfnamefont
  {R.}~\bibnamefont {Valentí}}, \bibinfo {author} {\bibfnamefont
  {G.}~\bibnamefont {Sangiovanni}},\ and\ \bibinfo {author} {\bibfnamefont
  {T.}~\bibnamefont {Wehling}},\ }\href {https://arxiv.org/abs/2309.08529}
  {\bibinfo {title} {Dynamical correlations and order in magic-angle twisted
  bilayer graphene}} (\bibinfo {year} {2023}),\ \Eprint
  {https://arxiv.org/abs/2309.08529} {arXiv:2309.08529 [cond-mat.str-el]}
  \BibitemShut {NoStop}%
\bibitem [{\citenamefont {Bennett}\ \emph {et~al.}(2024)\citenamefont
  {Bennett}, \citenamefont {Larson}, \citenamefont {Sharma}, \citenamefont
  {Carr},\ and\ \citenamefont {Kaxiras}}]{bennett2024twisted}%
  \BibitemOpen
  \bibfield  {author} {\bibinfo {author} {\bibfnamefont {D.}~\bibnamefont
  {Bennett}}, \bibinfo {author} {\bibfnamefont {D.~T.}\ \bibnamefont {Larson}},
  \bibinfo {author} {\bibfnamefont {L.}~\bibnamefont {Sharma}}, \bibinfo
  {author} {\bibfnamefont {S.}~\bibnamefont {Carr}},\ and\ \bibinfo {author}
  {\bibfnamefont {E.}~\bibnamefont {Kaxiras}},\ }\bibfield  {title} {\bibinfo
  {title} {Twisted bilayer graphene revisited: Minimal two-band model for
  low-energy bands},\ }\href {https://doi.org/10.1103/PhysRevB.109.155422}
  {\bibfield  {journal} {\bibinfo  {journal} {Phys. Rev. B}\ }\textbf {\bibinfo
  {volume} {109}},\ \bibinfo {pages} {155422} (\bibinfo {year}
  {2024})}\BibitemShut {NoStop}%
\bibitem [{Note1()}]{Note1}%
  \BibitemOpen
  \bibinfo {note} {In principle, the construction works for all supercells that
  are the result of a linear integer map applied to the primitive lattice
  vectors, i.e., $(\protect \bm {A}_1, \protect \bm {A}_2)^T = \protect
  \mathcal N\cdot (\protect \bm {a}_1, \protect \bm {a}_2)^T$, with $\protect
  \mathcal N$ a $2\times 2$ matrix with integer entries.}\BibitemShut {Stop}%
\bibitem [{SM()}]{SM}%
  \BibitemOpen
  \href@noop {} {\bibinfo {title} {{See Supplementary Material at [URL will be
  inserted by publisher] for details on the construction of supercell Wannier
  Hamiltonians, supercell Wannierization via single-shot projection, supercell
  Wannier functions for (i) Bernal bilayer graphene, (ii) monolayer graphene,
  (iii) rhombohedral (ABC) trilayer graphene, dual-gated Ohno-Coulomb
  interaction, and valley as a quantum number; including Refs.~\cite{%
  Wehling2011, aryasetiawan2004frequency, carr2019wannier,
  ghazaryan2023multilayer, jung2014accurate, kang2020topological,
  kinza2015low-energy, koepernik2023symmetry, marzari1997maximally,
  ramires2018electrically, seiler2022quantum, seiler2023interactiondriven,
  souza2001maximally, throckmorton2012fermions, wannier90, zhang2010bandsABC%
  }.}}}\BibitemShut {Stop}%
\bibitem [{\citenamefont {Haldane}(1988)}]{haldane1988model}%
  \BibitemOpen
  \bibfield  {author} {\bibinfo {author} {\bibfnamefont {F.~D.~M.}\
  \bibnamefont {Haldane}},\ }\bibfield  {title} {\bibinfo {title} {Model for a
  quantum hall effect without landau levels: Condensed-matter realization of
  the" parity anomaly"},\ }\href@noop {} {\bibfield  {journal} {\bibinfo
  {journal} {Physical Review Letters}\ }\textbf {\bibinfo {volume} {61}},\
  \bibinfo {pages} {2015} (\bibinfo {year} {1988})}\BibitemShut {NoStop}%
\bibitem [{\citenamefont {Ramires}\ and\ \citenamefont
  {Lado}(2018)}]{ramires2018electrically}%
  \BibitemOpen
  \bibfield  {author} {\bibinfo {author} {\bibfnamefont {A.}~\bibnamefont
  {Ramires}}\ and\ \bibinfo {author} {\bibfnamefont {J.~L.}\ \bibnamefont
  {Lado}},\ }\bibfield  {title} {\bibinfo {title} {Electrically tunable gauge
  fields in tiny-angle twisted bilayer graphene},\ }\href@noop {} {\bibfield
  {journal} {\bibinfo  {journal} {Physical review letters}\ }\textbf {\bibinfo
  {volume} {121}},\ \bibinfo {pages} {146801} (\bibinfo {year}
  {2018})}\BibitemShut {NoStop}%
\bibitem [{\citenamefont {Marzari}\ and\ \citenamefont
  {Vanderbilt}(1997)}]{marzari1997maximally}%
  \BibitemOpen
  \bibfield  {author} {\bibinfo {author} {\bibfnamefont {N.}~\bibnamefont
  {Marzari}}\ and\ \bibinfo {author} {\bibfnamefont {D.}~\bibnamefont
  {Vanderbilt}},\ }\bibfield  {title} {\bibinfo {title} {Maximally localized
  generalized wannier functions for composite energy bands},\ }\href@noop {}
  {\bibfield  {journal} {\bibinfo  {journal} {Physical review B}\ }\textbf
  {\bibinfo {volume} {56}},\ \bibinfo {pages} {12847} (\bibinfo {year}
  {1997})}\BibitemShut {NoStop}%
\bibitem [{\citenamefont {Kang}\ and\ \citenamefont
  {Vafek}(2019)}]{Kang2019strong}%
  \BibitemOpen
  \bibfield  {author} {\bibinfo {author} {\bibfnamefont {J.}~\bibnamefont
  {Kang}}\ and\ \bibinfo {author} {\bibfnamefont {O.}~\bibnamefont {Vafek}},\
  }\bibfield  {title} {\bibinfo {title} {Strong coupling phases of partially
  filled twisted bilayer graphene narrow bands},\ }\href@noop {} {\bibfield
  {journal} {\bibinfo  {journal} {Phys. Rev. Lett.}\ }\textbf {\bibinfo
  {volume} {122}},\ \bibinfo {pages} {246401} (\bibinfo {year}
  {2019})}\BibitemShut {NoStop}%
\bibitem [{\citenamefont {McCann}\ and\ \citenamefont
  {Fal’ko}(2006)}]{mccann2006landau}%
  \BibitemOpen
  \bibfield  {author} {\bibinfo {author} {\bibfnamefont {E.}~\bibnamefont
  {McCann}}\ and\ \bibinfo {author} {\bibfnamefont {V.~I.}\ \bibnamefont
  {Fal’ko}},\ }\bibfield  {title} {\bibinfo {title} {Landau-level degeneracy
  and quantum hall effect in a graphite bilayer},\ }\href@noop {} {\bibfield
  {journal} {\bibinfo  {journal} {Physical review letters}\ }\textbf {\bibinfo
  {volume} {96}},\ \bibinfo {pages} {086805} (\bibinfo {year}
  {2006})}\BibitemShut {NoStop}%
\bibitem [{\citenamefont {Zhang}\ \emph {et~al.}(2013)\citenamefont {Zhang},
  \citenamefont {MacDonald},\ and\ \citenamefont {Mele}}]{zhang2013valley}%
  \BibitemOpen
  \bibfield  {author} {\bibinfo {author} {\bibfnamefont {F.}~\bibnamefont
  {Zhang}}, \bibinfo {author} {\bibfnamefont {A.~H.}\ \bibnamefont
  {MacDonald}},\ and\ \bibinfo {author} {\bibfnamefont {E.~J.}\ \bibnamefont
  {Mele}},\ }\bibfield  {title} {\bibinfo {title} {Valley chern numbers and
  boundary modes in gapped bilayer graphene},\ }\href@noop {} {\bibfield
  {journal} {\bibinfo  {journal} {Proceedings of the National Academy of
  Sciences}\ }\textbf {\bibinfo {volume} {110}},\ \bibinfo {pages} {10546}
  (\bibinfo {year} {2013})}\BibitemShut {NoStop}%
\bibitem [{\citenamefont {Herzog-Arbeitman}\ \emph
  {et~al.}(2024{\natexlab{a}})\citenamefont {Herzog-Arbeitman}, \citenamefont
  {Wang}, \citenamefont {Liu}, \citenamefont {Tam}, \citenamefont {Qi},
  \citenamefont {Jia}, \citenamefont {Efetov}, \citenamefont {Vafek},
  \citenamefont {Regnault}, \citenamefont {Weng}, \citenamefont {Wu},
  \citenamefont {Bernevig},\ and\ \citenamefont
  {Yu}}]{herzogarbeitman2023moire}%
  \BibitemOpen
  \bibfield  {author} {\bibinfo {author} {\bibfnamefont {J.}~\bibnamefont
  {Herzog-Arbeitman}}, \bibinfo {author} {\bibfnamefont {Y.}~\bibnamefont
  {Wang}}, \bibinfo {author} {\bibfnamefont {J.}~\bibnamefont {Liu}}, \bibinfo
  {author} {\bibfnamefont {P.~M.}\ \bibnamefont {Tam}}, \bibinfo {author}
  {\bibfnamefont {Z.}~\bibnamefont {Qi}}, \bibinfo {author} {\bibfnamefont
  {Y.}~\bibnamefont {Jia}}, \bibinfo {author} {\bibfnamefont {D.~K.}\
  \bibnamefont {Efetov}}, \bibinfo {author} {\bibfnamefont {O.}~\bibnamefont
  {Vafek}}, \bibinfo {author} {\bibfnamefont {N.}~\bibnamefont {Regnault}},
  \bibinfo {author} {\bibfnamefont {H.}~\bibnamefont {Weng}}, \bibinfo {author}
  {\bibfnamefont {Q.}~\bibnamefont {Wu}}, \bibinfo {author} {\bibfnamefont
  {B.~A.}\ \bibnamefont {Bernevig}},\ and\ \bibinfo {author} {\bibfnamefont
  {J.}~\bibnamefont {Yu}},\ }\bibfield  {title} {\bibinfo {title} {Moir\'e
  fractional chern insulators. ii. first-principles calculations and continuum
  models of rhombohedral graphene superlattices},\ }\href
  {https://doi.org/10.1103/PhysRevB.109.205122} {\bibfield  {journal} {\bibinfo
   {journal} {Phys. Rev. B}\ }\textbf {\bibinfo {volume} {109}},\ \bibinfo
  {pages} {205122} (\bibinfo {year} {2024}{\natexlab{a}})}\BibitemShut
  {NoStop}%
\bibitem [{\citenamefont {Herzog-Arbeitman}\ \emph
  {et~al.}(2024{\natexlab{b}})\citenamefont {Herzog-Arbeitman}, \citenamefont
  {Yu}, \citenamefont {Călugăru}, \citenamefont {Hu}, \citenamefont
  {Regnault}, \citenamefont {Liu}, \citenamefont {Sarma}, \citenamefont
  {Vafek}, \citenamefont {Coleman}, \citenamefont {Tsvelik}, \citenamefont
  {da~Song},\ and\ \citenamefont {Bernevig}}]{herzogarbeitman2024topological}%
  \BibitemOpen
  \bibfield  {author} {\bibinfo {author} {\bibfnamefont {J.}~\bibnamefont
  {Herzog-Arbeitman}}, \bibinfo {author} {\bibfnamefont {J.}~\bibnamefont
  {Yu}}, \bibinfo {author} {\bibfnamefont {D.}~\bibnamefont {Călugăru}},
  \bibinfo {author} {\bibfnamefont {H.}~\bibnamefont {Hu}}, \bibinfo {author}
  {\bibfnamefont {N.}~\bibnamefont {Regnault}}, \bibinfo {author}
  {\bibfnamefont {C.}~\bibnamefont {Liu}}, \bibinfo {author} {\bibfnamefont
  {S.~D.}\ \bibnamefont {Sarma}}, \bibinfo {author} {\bibfnamefont
  {O.}~\bibnamefont {Vafek}}, \bibinfo {author} {\bibfnamefont
  {P.}~\bibnamefont {Coleman}}, \bibinfo {author} {\bibfnamefont
  {A.}~\bibnamefont {Tsvelik}}, \bibinfo {author} {\bibfnamefont
  {Z.}~\bibnamefont {da~Song}},\ and\ \bibinfo {author} {\bibfnamefont {B.~A.}\
  \bibnamefont {Bernevig}},\ }\href@noop {} {\bibinfo {title} {Topological
  heavy fermion principle for flat (narrow) bands with concentrated quantum
  geometry}} (\bibinfo {year} {2024}{\natexlab{b}}),\ \Eprint
  {https://arxiv.org/abs/2404.07253} {arXiv:2404.07253 [cond-mat.str-el]}
  \BibitemShut {NoStop}%
\bibitem [{\citenamefont {Soejima}\ \emph {et~al.}(2024)\citenamefont
  {Soejima}, \citenamefont {Dong}, \citenamefont {Wang}, \citenamefont {Wang},
  \citenamefont {Zaletel}, \citenamefont {Vishwanath},\ and\ \citenamefont
  {Parker}}]{soejima2024anomalous}%
  \BibitemOpen
  \bibfield  {author} {\bibinfo {author} {\bibfnamefont {T.}~\bibnamefont
  {Soejima}}, \bibinfo {author} {\bibfnamefont {J.}~\bibnamefont {Dong}},
  \bibinfo {author} {\bibfnamefont {T.}~\bibnamefont {Wang}}, \bibinfo {author}
  {\bibfnamefont {T.}~\bibnamefont {Wang}}, \bibinfo {author} {\bibfnamefont
  {M.~P.}\ \bibnamefont {Zaletel}}, \bibinfo {author} {\bibfnamefont
  {A.}~\bibnamefont {Vishwanath}},\ and\ \bibinfo {author} {\bibfnamefont
  {D.~E.}\ \bibnamefont {Parker}},\ }\href@noop {} {\bibinfo {title} {Anomalous
  hall crystals in rhombohedral multilayer graphene ii: General mechanism and a
  minimal model}} (\bibinfo {year} {2024}),\ \Eprint
  {https://arxiv.org/abs/2403.05522} {arXiv:2403.05522 [cond-mat.str-el]}
  \BibitemShut {NoStop}%
\bibitem [{\citenamefont {Zeng}\ \emph {et~al.}(2024)\citenamefont {Zeng},
  \citenamefont {Wolf}, \citenamefont {Huang}, \citenamefont {Wei},
  \citenamefont {Ghorashi}, \citenamefont {MacDonald},\ and\ \citenamefont
  {Cano}}]{zeng2024gatetunable}%
  \BibitemOpen
  \bibfield  {author} {\bibinfo {author} {\bibfnamefont {Y.}~\bibnamefont
  {Zeng}}, \bibinfo {author} {\bibfnamefont {T.~M.~R.}\ \bibnamefont {Wolf}},
  \bibinfo {author} {\bibfnamefont {C.}~\bibnamefont {Huang}}, \bibinfo
  {author} {\bibfnamefont {N.}~\bibnamefont {Wei}}, \bibinfo {author}
  {\bibfnamefont {S.~A.~A.}\ \bibnamefont {Ghorashi}}, \bibinfo {author}
  {\bibfnamefont {A.~H.}\ \bibnamefont {MacDonald}},\ and\ \bibinfo {author}
  {\bibfnamefont {J.}~\bibnamefont {Cano}},\ }\href@noop {} {\bibinfo {title}
  {Gate-tunable topological phases in superlattice modulated bilayer graphene}}
  (\bibinfo {year} {2024}),\ \Eprint {https://arxiv.org/abs/2401.04321}
  {arXiv:2401.04321 [cond-mat.mes-hall]} \BibitemShut {NoStop}%
\bibitem [{Note2()}]{Note2}%
  \BibitemOpen
  \bibinfo {note} {Using valley as a quantum number in the supercell model
  amounts to convoluting the $\protect \bm {q}=0$ valley operator with a
  $\protect \bm {q}=\protect \bm {G}_s$ charge density wave, see SM~\cite
  {SM}.}\BibitemShut {Stop}%
\bibitem [{\citenamefont {Profe}\ \emph {et~al.}(2024)\citenamefont {Profe},
  \citenamefont {Kennes},\ and\ \citenamefont {Klebl}}]{diverge}%
  \BibitemOpen
  \bibfield  {author} {\bibinfo {author} {\bibfnamefont {J.~B.}\ \bibnamefont
  {Profe}}, \bibinfo {author} {\bibfnamefont {D.~M.}\ \bibnamefont {Kennes}},\
  and\ \bibinfo {author} {\bibfnamefont {L.}~\bibnamefont {Klebl}},\ }\bibfield
   {title} {\bibinfo {title} {{divERGe implements various Exact Renormalization
  Group examples}},\ }\href {https://doi.org/10.21468/SciPostPhysCodeb.26}
  {\bibfield  {journal} {\bibinfo  {journal} {SciPost Phys. Codebases}\ ,\
  \bibinfo {pages} {26}} (\bibinfo {year} {2024})}\BibitemShut {NoStop}%
\bibitem [{\citenamefont {Aryasetiawan}\ \emph {et~al.}(2004)\citenamefont
  {Aryasetiawan}, \citenamefont {Imada}, \citenamefont {Georges}, \citenamefont
  {Kotliar}, \citenamefont {Biermann},\ and\ \citenamefont
  {Lichtenstein}}]{aryasetiawan2004frequency}%
  \BibitemOpen
  \bibfield  {author} {\bibinfo {author} {\bibfnamefont {F.}~\bibnamefont
  {Aryasetiawan}}, \bibinfo {author} {\bibfnamefont {M.}~\bibnamefont {Imada}},
  \bibinfo {author} {\bibfnamefont {A.}~\bibnamefont {Georges}}, \bibinfo
  {author} {\bibfnamefont {G.}~\bibnamefont {Kotliar}}, \bibinfo {author}
  {\bibfnamefont {S.}~\bibnamefont {Biermann}},\ and\ \bibinfo {author}
  {\bibfnamefont {A.}~\bibnamefont {Lichtenstein}},\ }\bibfield  {title}
  {\bibinfo {title} {Frequency-dependent local interactions and low-energy
  effective models from electronic structure calculations},\ }\href@noop {}
  {\bibfield  {journal} {\bibinfo  {journal} {Physical Review B}\ }\textbf
  {\bibinfo {volume} {70}},\ \bibinfo {pages} {195104} (\bibinfo {year}
  {2004})}\BibitemShut {NoStop}%
\bibitem [{\citenamefont {Kinza}\ and\ \citenamefont
  {Honerkamp}(2015)}]{kinza2015low-energy}%
  \BibitemOpen
  \bibfield  {author} {\bibinfo {author} {\bibfnamefont {M.}~\bibnamefont
  {Kinza}}\ and\ \bibinfo {author} {\bibfnamefont {C.}~\bibnamefont
  {Honerkamp}},\ }\bibfield  {title} {\bibinfo {title} {Low-energy effective
  interactions beyond the constrained random-phase approximation by the
  functional renormalization group},\ }\href
  {https://doi.org/10.1103/PhysRevB.92.045113} {\bibfield  {journal} {\bibinfo
  {journal} {Phys. Rev. B}\ }\textbf {\bibinfo {volume} {92}},\ \bibinfo
  {pages} {045113} (\bibinfo {year} {2015})}\BibitemShut {NoStop}%
\bibitem [{\citenamefont {Garcia-Ruiz}\ \emph {et~al.}(2023)\citenamefont
  {Garcia-Ruiz}, \citenamefont {Enaldiev}, \citenamefont {McEllistrim},\ and\
  \citenamefont {Fal’ko}}]{garcia2023mixed}%
  \BibitemOpen
  \bibfield  {author} {\bibinfo {author} {\bibfnamefont {A.}~\bibnamefont
  {Garcia-Ruiz}}, \bibinfo {author} {\bibfnamefont {V.}~\bibnamefont
  {Enaldiev}}, \bibinfo {author} {\bibfnamefont {A.}~\bibnamefont
  {McEllistrim}},\ and\ \bibinfo {author} {\bibfnamefont {V.~I.}\ \bibnamefont
  {Fal’ko}},\ }\bibfield  {title} {\bibinfo {title} {Mixed-stacking few-layer
  graphene as an elemental weak ferroelectric material},\ }\href@noop {}
  {\bibfield  {journal} {\bibinfo  {journal} {Nano Letters}\ }\textbf {\bibinfo
  {volume} {23}},\ \bibinfo {pages} {4120} (\bibinfo {year}
  {2023})}\BibitemShut {NoStop}%
\bibitem [{\citenamefont {McEllistrim}\ \emph {et~al.}(2023)\citenamefont
  {McEllistrim}, \citenamefont {Garcia-Ruiz}, \citenamefont {Goodwin},\ and\
  \citenamefont {Fal'ko}}]{mcellistrim2023spectroscopic}%
  \BibitemOpen
  \bibfield  {author} {\bibinfo {author} {\bibfnamefont {A.}~\bibnamefont
  {McEllistrim}}, \bibinfo {author} {\bibfnamefont {A.}~\bibnamefont
  {Garcia-Ruiz}}, \bibinfo {author} {\bibfnamefont {Z.~A.}\ \bibnamefont
  {Goodwin}},\ and\ \bibinfo {author} {\bibfnamefont {V.~I.}\ \bibnamefont
  {Fal'ko}},\ }\bibfield  {title} {\bibinfo {title} {Spectroscopic signatures
  of tetralayer graphene polytypes},\ }\href@noop {} {\bibfield  {journal}
  {\bibinfo  {journal} {Physical Review B}\ }\textbf {\bibinfo {volume}
  {107}},\ \bibinfo {pages} {155147} (\bibinfo {year} {2023})}\BibitemShut
  {NoStop}%
\bibitem [{\citenamefont {Kang}\ \emph {et~al.}(2020)\citenamefont {Kang},
  \citenamefont {Fang}, \citenamefont {Ye}, \citenamefont {Po}, \citenamefont
  {Denlinger}, \citenamefont {Jozwiak}, \citenamefont {Bostwick}, \citenamefont
  {Rotenberg}, \citenamefont {Kaxiras}, \citenamefont {Checkelsky},\ and\
  \citenamefont {Comin}}]{kang2020topological}%
  \BibitemOpen
  \bibfield  {author} {\bibinfo {author} {\bibfnamefont {M.}~\bibnamefont
  {Kang}}, \bibinfo {author} {\bibfnamefont {S.}~\bibnamefont {Fang}}, \bibinfo
  {author} {\bibfnamefont {L.}~\bibnamefont {Ye}}, \bibinfo {author}
  {\bibfnamefont {H.~C.}\ \bibnamefont {Po}}, \bibinfo {author} {\bibfnamefont
  {J.}~\bibnamefont {Denlinger}}, \bibinfo {author} {\bibfnamefont
  {C.}~\bibnamefont {Jozwiak}}, \bibinfo {author} {\bibfnamefont
  {A.}~\bibnamefont {Bostwick}}, \bibinfo {author} {\bibfnamefont
  {E.}~\bibnamefont {Rotenberg}}, \bibinfo {author} {\bibfnamefont
  {E.}~\bibnamefont {Kaxiras}}, \bibinfo {author} {\bibfnamefont {J.~G.}\
  \bibnamefont {Checkelsky}},\ and\ \bibinfo {author} {\bibfnamefont
  {R.}~\bibnamefont {Comin}},\ }\bibfield  {title} {\bibinfo {title}
  {Topological flat bands in frustrated kagome lattice {CoSn}},\ }\href
  {https://doi.org/10.1038/s41467-020-17465-1} {\bibfield  {journal} {\bibinfo
  {journal} {Nature Communications}\ }\textbf {\bibinfo {volume} {11}},\
  \bibinfo {pages} {4004} (\bibinfo {year} {2020})}\BibitemShut {NoStop}%
\bibitem [{\citenamefont {Koepernik}\ \emph {et~al.}(2023)\citenamefont
  {Koepernik}, \citenamefont {Janson}, \citenamefont {Sun},\ and\ \citenamefont
  {Van Den~Brink}}]{koepernik2023symmetry}%
  \BibitemOpen
  \bibfield  {author} {\bibinfo {author} {\bibfnamefont {K.}~\bibnamefont
  {Koepernik}}, \bibinfo {author} {\bibfnamefont {O.}~\bibnamefont {Janson}},
  \bibinfo {author} {\bibfnamefont {Y.}~\bibnamefont {Sun}},\ and\ \bibinfo
  {author} {\bibfnamefont {J.}~\bibnamefont {Van Den~Brink}},\ }\bibfield
  {title} {\bibinfo {title} {Symmetry-conserving maximally projected wannier
  functions},\ }\href@noop {} {\bibfield  {journal} {\bibinfo  {journal}
  {Physical Review B}\ }\textbf {\bibinfo {volume} {107}},\ \bibinfo {pages}
  {235135} (\bibinfo {year} {2023})}\BibitemShut {NoStop}%
\bibitem [{\citenamefont {Souza}\ \emph {et~al.}(2001)\citenamefont {Souza},
  \citenamefont {Marzari},\ and\ \citenamefont
  {Vanderbilt}}]{souza2001maximally}%
  \BibitemOpen
  \bibfield  {author} {\bibinfo {author} {\bibfnamefont {I.}~\bibnamefont
  {Souza}}, \bibinfo {author} {\bibfnamefont {N.}~\bibnamefont {Marzari}},\
  and\ \bibinfo {author} {\bibfnamefont {D.}~\bibnamefont {Vanderbilt}},\
  }\bibfield  {title} {\bibinfo {title} {Maximally localized wannier functions
  for entangled energy bands},\ }\href
  {https://doi.org/10.1103/PhysRevB.65.035109} {\bibfield  {journal} {\bibinfo
  {journal} {Phys. Rev. B}\ }\textbf {\bibinfo {volume} {65}},\ \bibinfo
  {pages} {035109} (\bibinfo {year} {2001})}\BibitemShut {NoStop}%
\bibitem [{\citenamefont {Throckmorton}\ and\ \citenamefont
  {Vafek}(2012)}]{throckmorton2012fermions}%
  \BibitemOpen
  \bibfield  {author} {\bibinfo {author} {\bibfnamefont {R.~E.}\ \bibnamefont
  {Throckmorton}}\ and\ \bibinfo {author} {\bibfnamefont {O.}~\bibnamefont
  {Vafek}},\ }\bibfield  {title} {\bibinfo {title} {Fermions on bilayer
  graphene: Symmetry breaking for b= 0 and $\nu$= 0},\ }\href@noop {}
  {\bibfield  {journal} {\bibinfo  {journal} {Physical Review B}\ }\textbf
  {\bibinfo {volume} {86}},\ \bibinfo {pages} {115447} (\bibinfo {year}
  {2012})}\BibitemShut {NoStop}%
\bibitem [{\citenamefont {Mostofi}\ \emph {et~al.}(2008)\citenamefont
  {Mostofi}, \citenamefont {Yates}, \citenamefont {Lee}, \citenamefont {Souza},
  \citenamefont {Vanderbilt},\ and\ \citenamefont {Marzari}}]{wannier90}%
  \BibitemOpen
  \bibfield  {author} {\bibinfo {author} {\bibfnamefont {A.~A.}\ \bibnamefont
  {Mostofi}}, \bibinfo {author} {\bibfnamefont {J.~R.}\ \bibnamefont {Yates}},
  \bibinfo {author} {\bibfnamefont {Y.-S.}\ \bibnamefont {Lee}}, \bibinfo
  {author} {\bibfnamefont {I.}~\bibnamefont {Souza}}, \bibinfo {author}
  {\bibfnamefont {D.}~\bibnamefont {Vanderbilt}},\ and\ \bibinfo {author}
  {\bibfnamefont {N.}~\bibnamefont {Marzari}},\ }\bibfield  {title} {\bibinfo
  {title} {wannier90: A tool for obtaining maximally-localised wannier
  functions},\ }\href@noop {} {\bibfield  {journal} {\bibinfo  {journal}
  {Computer physics communications}\ }\textbf {\bibinfo {volume} {178}},\
  \bibinfo {pages} {685} (\bibinfo {year} {2008})}\BibitemShut {NoStop}%
\bibitem [{\citenamefont {Zhang}\ \emph {et~al.}(2010)\citenamefont {Zhang},
  \citenamefont {Sahu}, \citenamefont {Min},\ and\ \citenamefont
  {MacDonald}}]{zhang2010bandsABC}%
  \BibitemOpen
  \bibfield  {author} {\bibinfo {author} {\bibfnamefont {F.}~\bibnamefont
  {Zhang}}, \bibinfo {author} {\bibfnamefont {B.}~\bibnamefont {Sahu}},
  \bibinfo {author} {\bibfnamefont {H.}~\bibnamefont {Min}},\ and\ \bibinfo
  {author} {\bibfnamefont {A.~H.}\ \bibnamefont {MacDonald}},\ }\bibfield
  {title} {\bibinfo {title} {Band structure of $abc$-stacked graphene
  trilayers},\ }\href {https://doi.org/10.1103/PhysRevB.82.035409} {\bibfield
  {journal} {\bibinfo  {journal} {Phys. Rev. B}\ }\textbf {\bibinfo {volume}
  {82}},\ \bibinfo {pages} {035409} (\bibinfo {year} {2010})}\BibitemShut
  {NoStop}%
\end{thebibliography}%

\clearpage
\arxivSubmit

\end{document}